\documentclass[fleqn,a4paper]{article}

\usepackage[utf8]{inputenc} % why not type "Bézout" with unicode?
\usepackage[T1]{fontenc} % vector fonts plz
\usepackage{fullpage,amsmath,amssymb,latexsym,graphicx}
\usepackage{newtxtext,newtxmath} % Times for PR
\usepackage{appendix}
\usepackage[dvipsnames]{xcolor}
\usepackage[
  colorlinks=true,
  urlcolor=MidnightBlue,
  citecolor=MidnightBlue,
  filecolor=MidnightBlue,
  linkcolor=MidnightBlue
]{hyperref} % ref and cite links with pretty colors
\usepackage[
  style=phys,
  eprint=true,
  maxnames = 100
]{biblatex}
\usepackage{anyfontsize,authblk}

\begin{document}

\title{
  How to count in hierarchical landscapes:\\ a `full' solution to mean-field complexity
}

\author{Jaron Kent-Dobias}
\author{Jorge Kurchan}
\affil{Laboratoire de Physique de l'Ecole Normale Supérieure, Paris, France}

\maketitle
\begin{abstract}
  We derive the general solution for counting the stationary points of
  mean-field complex landscapes. It incorporates Parisi's solution
  for the ground state, as it should. Using this solution, we count the
  stationary points of two models: one with multi-step replica symmetry
  breaking, and one with full replica symmetry breaking.
\end{abstract}

\tableofcontents

\section{Introduction}

The computation of the number of metastable states of mean field spin glasses
goes back to the beginning of the field. Over forty years ago, Bray and Moore
\cite{Bray_1980_Metastable} attempted the first calculation for the
Sherrington--Kirkpatrick model, in a paper remarkable for being one of the
first applications of a replica symmetry breaking (RSB) scheme. As was clear
when the actual ground-state of the model was computed by Parisi with a
different scheme, the Bray--Moore result was not exact, and the problem has
been open ever since \cite{Parisi_1979_Infinite}.  To date, the program of
computing the number of stationary points---minima, saddle points, and
maxima---of  mean-field complex landscapes has been only carried out for a small subset of
models, including most notably the (pure) $p$-spin model ($p>2$)
\cite{Rieger_1992_The, Crisanti_1995_Thouless-Anderson-Palmer, Cavagna_1997_An, Cavagna_1998_Stationary} and for similar
energy functions inspired by molecular biology, evolution, and machine learning
\cite{Maillard_2020_Landscape, Ros_2019_Complex, Altieri_2021_Properties}.  In
a parallel development, it has evolved into an active field of probability
theory \cite{Auffinger_2012_Random, Auffinger_2013_Complexity,
BenArous_2019_Geometry}.

In this paper we present what we argue is the general replica ansatz for the
number of stationary points of generic mean-field models, which we expect to
include the Sherrington--Kirkpatrick model. It reproduces the Parisi result in
the limit of small temperature for the lowest states, as it should.

To understand the importance of this computation, consider the following
situation. When one solves the problem of spheres in large dimensions, one
finds that there is a transition at a given temperature to a one-step replica symmetry
breaking (1RSB) phase at a Kauzmann temperature, and, at a lower temperature,
another  transition to  a full RSB (FRSB) phase (see \cite{Gross_1985_Mean-field,
Gardner_1985_Spin}, the so-called `Gardner' phase
\cite{Charbonneau_2014_Fractal}).  Now, this transition involves the lowest
equilibrium states. Because they are obviously unreachable at any reasonable
timescale, a common question is: what is the signature of the Gardner
transition line for higher than equilibrium energy-densities? This is a
question whose answers are significant to interpreting the results of myriad
experiments and simulations \cite{Xiao_2022_Probing, Hicks_2018_Gardner,
Liao_2019_Hierarchical, Dennis_2020_Jamming, Charbonneau_2015_Numerical,
Li_2021_Determining, Seguin_2016_Experimental, Geirhos_2018_Johari-Goldstein,
Hammond_2020_Experimental, Albert_2021_Searching} (see, for a review
\cite{Berthier_2019_Gardner}).  For example, when studying `jamming' at zero
temperature, the question is posed as `on what side of the 1RSB--FRSB
transition are high energy (or low density) states reachable dynamically?' One
approach to answering such questions makes use of `state following,'
which tracks metastable thermodynamic configurations to their zero temperature
limit \cite{Rainone_2015_Following, Biroli_2016_Breakdown,
Rainone_2016_Following, Biroli_2018_Liu-Nagel, Urbani_2017_Shear}.  In the
present paper we give a purely geometric appoarch: we consider the local energy
minima at a given energy and study their number and other properties: the
solution involves a replica-symmetry breaking scheme that is well-defined, and
corresponds directly to the topological characteristics of those minima.

Perhaps the most interesting application of this computation is in the context
of optimization problems, see for example \cite{Gamarnik_2021_The,
ElAlaoui_2022_Sampling, Huang_2021_Tight}. A question that appears there is how
to define a `threshold' level, the lowest energy level that good algorithms can
expect to reach. This notion was introduced in the context of the pure $p$-spin
models, as the energy at which level sets of the energy in phase-space
percolate, explaining why dynamics never go below that level
\cite{Cugliandolo_1993_Analytical}.  The notion of a `threshold' for more
complicated landscapes has later been invoked several times, never to our
knowledge in a clear and unambiguous way. One of the purposes of this paper is
to give a sufficiently detailed characterization of a general landscape so that
a meaningful general notion of threshold may  be introduced -- if this is at all
possible.

The format of this paper is as follows. In \S\ref{sec:model}, we introduce the
mean-field model of study, the mixed $p$-spin spherical model. In
\S\ref{sec:equilibrium} we review details of the equilibrium solution that are
relevant to our study of the landscape complexity. In \S\ref{sec:complexity} we
derive a generic form for the complexity. In \S\ref{sec:ansatz} we make and
review the hierarchical replica symmetry breaking ansatz used to solve the
complexity. In \S\ref{sec:supersymmetric} we write down the solution in a
specific and limited regime, which is nonetheless helpful as it gives a
foothold for numerically computing the complexity everywhere else.
\S\ref{sec:frsb} explains aspects of the solution specific to the case of full
RSB, and derives the replica symmetric to full FRSB (RS--FRSB) transition line.
\S\ref{sec:examples} details the landscape topology of two example models: a
$3+16$ model with a 2RSB ground state and a 1RSB complexity, and a $2+4$ with a
FRSB ground state and a FRSB complexity. Finally \S\ref{sec:interpretation}
provides some interpretation of our results.

\section{The model}
\label{sec:model}

For definiteness, we consider the mixed $p$-spin spherical model, whose Hamiltonian
\begin{equation} \label{eq:hamiltonian}
  H(\mathbf s)=-\sum_p\frac1{p!}\sum_{i_1\cdots i_p}^NJ^{(p)}_{i_1\cdots i_p}s_{i_1}\cdots s_{i_p}
\end{equation}
is defined for vectors $\mathbf s\in\mathbb R^N$ confined to the sphere
$\|\mathbf s\|^2=N$.  The coupling coefficients $J$ are taken at random, with
zero mean and variance $\overline{(J^{(p)})^2}=a_pp!/2N^{p-1}$ chosen so that
the energy is typically extensive. The overbar will always denote an average
over the coefficients $J$. The factors $a_p$ in the variances are freely chosen
constants that define the particular model. For instance, the so-called `pure'
models have $a_p=1$ for some $p$ and all others zero.

The variance of the couplings implies that the covariance of the energy with
itself depends on only the dot product (or overlap) between two configurations.
In particular, one finds
\begin{equation} \label{eq:covariance}
  \overline{H(\mathbf s_1)H(\mathbf s_2)}=Nf\left(\frac{\mathbf s_1\cdot\mathbf s_2}N\right)
\end{equation}
where $f$ is defined by the series
\begin{equation}
  f(q)=\frac12\sum_pa_pq^p
\end{equation}
One needn't start with a Hamiltonian like
\eqref{eq:hamiltonian}, defined as a series: instead, the covariance rule
\eqref{eq:covariance} can be specified for arbitrary, non-polynomial $f$, as in
the `toy model' of M\'ezard and Parisi \cite{Mezard_1992_Manifolds}.

The family of mixed $p$-spin models may be considered as the most general
models of generic Gaussian functions on the sphere.  To constrain the model to
the sphere, we use a Lagrange multiplier $\mu$, with the total energy being
\begin{equation}
  H(\mathbf s)+\frac\mu2(\|\mathbf s\|^2-N)
\end{equation}
For reasons that will become clear in \S\ref{subsec:hess}, we refer to $\mu$
as the \emph{stability parameter}.  At any stationary point, the gradient and
Hessian are given by
\begin{align}
  \nabla H(\mathbf s,\mu)=\partial H(\mathbf s)+\mu\mathbf s &&
  \operatorname{Hess}H(\mathbf s,\mu)=\partial\partial H(\mathbf s)+\mu I
\end{align}
where $\partial=\frac\partial{\partial\mathbf s}$ always. An important
observation was made by Bray and Dean  \cite{Bray_2007_Statistics} that
gradient and Hessian are independent for Gaussian random functions.  The
average over disorder breaks into a product of two independent averages, one
for any function of the gradient and one for any function of the Hessian.  In
particular, the number of negative eigenvalues at a stationary point, which
sets the index $\mathcal I$ of the saddle, is a function of the Hessian alone
(see Fyodorov \cite{Fyodorov_2007_Replica} for a detailed discussion).

\section{Equilibrium}
\label{sec:equilibrium}

Here we review the equilibrium solution, which has been studied in detail
\cite{Crisanti_1992_The, Crisanti_1993_The, Crisanti_2004_Spherical,
Crisanti_2006_Spherical}. For a succinct review, see \cite{Folena_2020_The}.
The free energy, averaged over disorder, is
\begin{equation}
  \beta F = - \overline{\ln \int d\mathbf s \;\delta(\|\mathbf s\|^2-N)\, e^{-\beta H(\mathbf s)}}
\end{equation}
Once $n$ replicas are introduced to treat the logarithm, the fields $\mathbf
s_a$ can be replaced with the new $n\times n$ matrix field $Q_{ab}\equiv(\mathbf
s_a\cdot\mathbf s_b)/N$. This yields for the free energy
\begin{equation} \label{eq:eq.free.energy}
  \beta F=-1-\ln2\pi-\frac12\lim_{n\to0}\frac1n\left(\beta^2\sum_{ab}^nf(Q_{ab})+\ln\det Q\right)
\end{equation}
which must be evaluated at the $Q$ which maximizes this expression and whose
diagonal is one. The solution is generally a hierarchical matrix \emph{à la}
Parisi. The properties of these matrices is reviewed in \S\ref{sec:dict},
including how to write down \eqref{eq:eq.free.energy} in terms of their
parameters.

The free energy can also be written in a functional form, which is necessary
for working with the solution in the limit $k\to\infty$, the so-called full
replica symmetry breaking (FRSB). If $P(q)$ is the probability distribution for
elements $q$ in a row of the matrix, then define $\chi(q)$ by
\begin{equation}
  \chi(q)=\int_q^1dq'\,\int_0^{q'}dq''\,P(q'')
\end{equation}
Since it is the double integral of a probability distribution, $\chi$ must be
concave, monotonically decreasing, and have $\chi(1)=0$ and $\chi'(1)=-1$. The
function $\chi$ turns out to have an interpretation as the spectrum of the
hierarchical matrix $Q$.  Using standard arguments, the free energy can be
written as a functional over $\chi$ as
\begin{equation}
  \beta F=-1-\ln2\pi-\frac12\int_0^1dq\,\left(\beta^2f''(q)\chi(q)+\frac1{\chi(q)}\right)
\end{equation}
which must be maximized with respect to $\chi$ given the constraints outlined
above.

In our study of the landscape, the free energy will not be directly relevant
anywhere except at the ground state, when the temperature is zero or
$\beta\to\infty$. Here, the measure will be concentrated in the lowest minima,
and the average energy $\langle
E\rangle_0=\lim_{\beta\to\infty}\frac\partial{\partial\beta}\beta F$ will
correspond to the ground state energy $E_0$.  The zero temperature limit is
most easily obtained by putting $x_i=\tilde x_ix_k$ and
$x_k=\tilde\beta/\beta$, $q_k=1-z/\beta$, which ensures the $\tilde x_i$,
$\tilde\beta$, and $z$ have nontrivial limits. Inserting the ansatz and taking
the limit, carefully treating the $k$th term in each sum separately from the
rest, one can show after some algebra that
\begin{equation} \label{eq:ground.state.free.energy}
  \tilde\beta\langle E\rangle_0=\tilde\beta\lim_{\beta\to\infty}\frac{\partial(\beta F)}{\partial\beta}
  =-\frac12z\tilde\beta f'(1)-\frac12\lim_{n\to0}\frac1n\left(
    \tilde\beta^2\sum_{ab}^nf(\tilde Q_{ab})+\ln\det(\tilde\beta z^{-1}\tilde Q+I)
  \right)
\end{equation}
where $\tilde Q$ is a $(k-1)$RSB matrix with entries $\tilde q_1=\lim_{\beta\to\infty}q_1$, \dots, $\tilde q_{k-1}=\lim_{\beta\to\infty}q_{k-1}$ parameterized by $\tilde x_1,\ldots,\tilde x_{k-1}$.
This is a ($k-1$)RSB ansatz whose spectrum in the determinant is scaled by
$\tilde\beta z^{-1}$ and shifted by 1, with effective temperature
$\tilde\beta$, and an extra term. In the continuum case, this is
\begin{equation} \label{eq:ground.state.free.energy.cont}
  \tilde\beta \langle E\rangle_0
  =-\frac12z\tilde\beta f'(1)-\frac12\int_0^1 dq\left(
    \tilde\beta^2f''(q)\tilde\chi(q)+\frac1{\tilde\chi(q)+\tilde\beta z^{-1}}
  \right)
\end{equation}
where $\tilde\chi$ is bound by the same constraints as $\chi$.

The zero temperature limit of the free energy loses one level of replica
symmetry breaking. Physically, this is a result of the fact that in $k$RSB,
$q_k$ gives the overlap within a state, i.e., within the basin of a well inside
the energy landscape. At zero temperature, the measure is completely localized
on the bottom of the well, and therefore the overlap within each state becomes
one. We will see that the complexity of low-energy stationary points in
Kac--Rice computation is also given by a $(k-1)$RSB anstaz. Heuristically, this is because
each stationary point also has no width and therefore overlap one with itself.

\section{Landscape complexity}
\label{sec:complexity}

The stationary points of a function can be counted using the Kac--Rice formula,
which integrates over the function's domain a $\delta$-function containing
the gradient multiplied by the absolute value of the determinant
\cite{Rice_1939_The, Kac_1943_On}. It gives the number of stationary points $\mathcal N$ as
\begin{equation} \label{eq:kac-rice}
  \mathcal N
    =\int d\mathbf s\, d\mu\,\delta\big(\tfrac12(\|\mathbf s\|^2-N)\big)\,\delta\big(\nabla H(\mathbf s,\mu)\big)\,\big|\det\operatorname{Hess}H(\mathbf s,\mu)\big|
\end{equation}
It is more interesting to count stationary points which share certain
properties, like energy density $E$ or index density $\mathcal I$. These properties can
be fixed by inserting additional $\delta$-functions into the integral. Rather
than fix the index directly, we fix the trace of the Hessian, which we'll soon
show is equivalent to fixing the value $\mu$, and fixing $\mu$ fixes the index
to within order one. Inserting these $\delta$-functions, we arrive at
\begin{equation}
  \begin{aligned}
    \mathcal N(E, \mu^*)
    &=\int d\mathbf s\, d\mu\,\delta\big(\tfrac12(\|\mathbf s\|^2-N)\big)\,\delta\big(\nabla H(\mathbf s,\mu)\big)\,\big|\det\operatorname{Hess}H(\mathbf s,\mu)\big| \\
    &\hspace{10pc}\times\delta\big(NE-H(\mathbf s)\big)\delta\big(N\mu^*-\operatorname{Tr}\operatorname{Hess}H(\mathbf s,\mu)\big)
  \end{aligned}
\end{equation}
This number will typically be exponential in $N$. In order to find the typical
count when disorder is averaged, we want to average its logarithm instead,
which is known as the complexity:
\begin{equation}
  \Sigma(E,\mu^*)=\lim_{N\to\infty}\frac1N\overline{\log\mathcal N(E, \mu^*})
\end{equation}
If one averages over $\mathcal N$ and afterward takes its logarithm, one arrives at the so-called {\em annealed} complexity
\begin{equation}
  \Sigma_\mathrm a(E,\mu^*)
  =\lim_{N\to\infty}\frac1N\log\overline{\mathcal N(E,\mu^*)}
\end{equation}
The annealed complexity has been previously computed for the mixed $p$-spin models
\cite{BenArous_2019_Geometry}.  The annealed complexity is known to equal the
actual (quenched) complexity in circumstances where there is at most one level
of replica symmetry breaking in the model's equilibrium. This is the case for
the pure $p$-spin models, or for mixed models where $1/\sqrt{f''(q)}$ is a
convex function. However, it fails dramatically for models with higher replica
symmetry breaking. For instance, when $f(q)=\frac12(q^2+\frac1{16}q^4)$ (a
model we study in detail later), the annealed complexity predicts that minima
vanish well before the dominant saddles, a contradiction for any bounded
function.

A sometimes more illuminating quantity is the Legendre transform $G$ of the complexity, defined by
\begin{equation}
    e^{NG(\hat \beta, \mu^*)} = \int dE \; e^{ -\hat \beta E +\Sigma(\hat \beta, \mu^*)}
\end{equation}
There will be a critical value $\hat \beta_c$ beyond  which the complexity is zero: above
this value the measure is split between the lowest $O(1)$ energy states. We shall not study here this  regime that interpolates between the dynamically relevant and the equilibrium states, but just mention that
it is an interesting object of study.

\subsection{The replicated problem}

The replicated Kac--Rice formula was introduced by Ros et
al.~\cite{Ros_2019_Complex}, and its effective action for the mixed $p$-spin
model has previously been computed by Folena et
al.~\cite{Folena_2020_Rethinking}. Here we review the derivation.

In order to average the complexity over disorder, we must deal with the
logarithm. We use the standard replica trick to convert the logarithm into a
product, which gives
\begin{equation}
  \begin{aligned}
    \log\mathcal N(E,\mu^*)
    &=\lim_{n\to0}\frac\partial{\partial n}\mathcal N^n(E,\mu^*) \\
    &=\lim_{n\to0}\frac\partial{\partial n}\int\prod_a^n d\mathbf s_a\,d\mu_a\,
    \delta\big(\tfrac12(\|\mathbf s_a\|^2-N)\big)\,\delta\big(\nabla H(\mathbf s_a,\mu_a)\big)\,\big|\det\operatorname{Hess}H(\mathbf s_a,\mu_a)\big| \\
    &\hspace{13pc}  \times\delta\big(NE-H(\mathbf s_a)\big)\delta\big(N\mu^*-\operatorname{Tr}\operatorname{Hess}H(\mathbf s_a,\mu_a)\big)
  \end{aligned}
\end{equation}
As discussed in \S\ref{sec:model}, it has been shown that to the largest order
in $N$, the Hessian of Gaussian random functions is independent from their
gradient, once both are conditioned on certain properties. Here, they are only
related by their shared value of $\mu$. Because of this statistical
independence, we may write
\begin{equation}
  \begin{aligned}
    \Sigma(E, \mu^*)
    &=\lim_{N\to\infty}\frac1N\lim_{n\to0}\frac\partial{\partial n}\int\left(\prod_a^nd\mathbf s_a\,d\mu_a\right)\,
    \overline{\prod_a^n \delta\big(\tfrac12(\|\mathbf s_a\|^2-N)\big)\,\delta\big(\nabla H(\mathbf s_a,\mu_a)\big)\delta(NE-H(\mathbf s_a))}\\
    &\hspace{10pc}
    \times
    \overline{\prod_a^n |\det\operatorname{Hess}(\mathbf s_a,\mu_a)|\,\delta\big(N\mu^*-\operatorname{Tr}\operatorname{Hess}H(\mathbf s_a,\mu_a)\big)}
  \end{aligned}
\end{equation}
which simplifies matters. The average of the two factors may now be treated separately.

\subsubsection{The Hessian factors}
\label{subsec:hess}

The spectrum of the matrix $\partial\partial H(\mathbf s)$ is uncorrelated from the
gradient. In the large-$N$ limit, for almost every point and realization of
disorder it is a GOE matrix with variance
\begin{equation}
  \overline{(\partial_i\partial_jH(\mathbf s))^2}=\frac1Nf''(1)\delta_{ij}
\end{equation}
Therefore in that limit its spectrum is given by the Wigner semicircle with radius $\sqrt{4f''(1)}$, or
\begin{equation}
  \rho(\lambda)=\begin{cases}
    \frac1{2\pi f''(1)}\sqrt{4f''(1)-\lambda^2} & \lambda^2\leq 4f''(1) \\
    0 & \text{otherwise}
  \end{cases}
\end{equation}
The spectrum of the Hessian $\operatorname{Hess}H(\mathbf s,\mu)$ is the same
semicircle shifted by $\mu$, or $\rho(\lambda+\mu)$.  The stability parameter
$\mu$ thus fixes the center of the spectrum of the Hessian. The semicircle radius
$\mu_m=\sqrt{4f''(1)}$ is a kind of threshold. When $\mu$ is taken to be within
the range $\pm\mu_m$, the critical points have index density
\begin{equation}
  \mathcal I(\mu)=\int_0^\infty d\lambda\,\rho(\lambda+\mu)
  =\frac12-\frac1\pi\left[
    \arctan\left(\frac\mu{\sqrt{\mu_m^2-\mu^2}}\right)
    +\frac\mu{\mu_m^2}\sqrt{\mu_m^2-\mu^2}
  \right]
\end{equation}
When $\mu>\mu_m$, the critical points are minima whose sloppiest eigenvalue is
$\mu-\mu_m$. When $\mu=\mu_m$, the critical points are marginal minima, with
flat directions in their spectrum. This property of $\mu$ is why we've named it
the stability parameter: it governs the stability of stationary points, and for
unstable ones it governs their index.

To largest order in $N$, the average over the product of determinants
factorizes into the product of averages, each of which is given by the same
expression depending only on $\mu$ \cite{Ros_2019_Complex}. We therefore find
\begin{equation}
  \overline{\prod_a^n |\det\operatorname{Hess}(\mathbf s_a,\mu_a)|\,\delta\big(N\mu^*-\operatorname{Tr}\operatorname{Hess}H(\mathbf s_a,\mu_a)\big)}
  \rightarrow \prod_a^ne^{N{\cal D}(\mu_a)}\delta\big(N(\mu^*-\mu_a)\big)
\end{equation}
where the function $\mathcal D$ is defined by
\begin{equation}
  \begin{aligned}
    \mathcal D(\mu)
    &=\frac1N\overline{\ln|\det\operatorname{Hess}H(s,\mu)|}
    =\int d\lambda\,\rho(\lambda+\mu)\ln|\lambda| \\
    &=\operatorname{Re}\left\{
    \frac12\left(1+\frac\mu{2f''(1)}\left(\mu-\sqrt{\mu^2-4f''(1)}\right)\right)
    -\ln\left(\frac1{2f''(1)}\left(\mu-\sqrt{\mu^2-4f''(1)}\right)\right)
  \right\}
  \end{aligned}
\end{equation}

By fixing the trace of the Hessian, we have effectively fixed
the value of the stability $\mu$ in all replicas to the value $\mu^*$.
\begin{itemize}
  \item For $\mu^*<\mu_m$, this amounts to fixing the index density. Since the
    overwhelming majority of saddles have a semicircle distribution, the
    fluctuations are rarer than exponential.
  \item For the gapped case $\mu^*>\mu_m$, there is an exponentially small
    probability that $r=1,2,...$ eigenvalues detach from the semicircle in such
    a way that the index is in fact $N {\cal{I}}=r$.  We shall not discuss
    these subextensive index fluctuations in this paper, the interested reader
    may find what is needed in \cite{Auffinger_2013_Complexity}.
\end{itemize}

\subsubsection{The gradient factors}

The $\delta$-functions in the remaining factor are treated by writing them in
the Fourier basis.  Introducing auxiliary fields $\hat{\mathbf s}_a$ and
$\hat\beta$ for this purpose, for each replica replica one writes
\begin{equation}
  \begin{aligned}
    &\delta\big(\tfrac12(\|\mathbf s_a\|^2-N)\big)\,\delta\big(\nabla H(\mathbf s_a,\mu^*)\big)\delta(NE-H(\mathbf s_a)) \\
    &\hspace{12pc}=\int\frac{d\hat\mu}{2\pi}\,\frac{d\hat\beta}{2\pi}\,\frac{d\hat{\mathbf s}_a}{(2\pi)^N}
      e^{\frac12\hat\mu(\|\mathbf s_a\|^2-N)+\hat\beta(NE-H(\mathbf s_a))+i\hat{\mathbf s}_a\cdot(\partial H(\mathbf s_a)+\mu^*\mathbf s_a)}
  \end{aligned}
\end{equation}
Anticipating a Parisi-style solution, we don't label $\hat\mu$ or $\hat\beta$
with replica indices, since replica vectors won't be broken in the scheme.  The
average over disorder can now be taken for the pieces which depend explicitly
on the Hamiltonian, and since everything is Gaussian this gives
\begin{equation}
  \begin{aligned}
    \overline{
      \exp\left[
        \sum_a^n(i\hat {\mathbf s}_a\cdot\partial_a-\hat\beta)H(s_a)
      \right]
    }
    &=\exp\left[
        \frac12\sum_{ab}^n
        (i\hat{\mathbf s}_a\cdot\partial_a-\hat\beta)
        (i\hat{\mathbf s}_b\cdot\partial_b-\hat\beta)
        \overline{H(\mathbf s_a)H(\mathbf s_b)}
      \right] \\
    &=\exp\left[
        \frac N2\sum_{ab}^n
        (i\hat{\mathbf s}_a\cdot\partial_a-\hat\beta)
        (i\hat{\mathbf s}_b\cdot\partial_b-\hat\beta)
        f\left(\frac{\mathbf s_a\cdot\mathbf s_b}N\right)
      \right] \\
    &\hspace{-14em}=\exp\left\{
        \frac N2\sum_{ab}^n
        \left[
          \hat\beta^2f\left(\frac{{\mathbf s}_a\cdot {\mathbf s}_b}N\right)
          -2i\hat\beta\frac{\hat {\mathbf s}_a\cdot {\mathbf s}_b}Nf'\left(\frac{{\mathbf s}_a\cdot {\mathbf s}_b}N\right)
          -\frac{\hat {\mathbf s}_a\cdot \hat {\mathbf s}_b}Nf'\left(\frac{{\mathbf s}_a\cdot {\mathbf s}_b}N\right)
          +\left(i\frac{\hat {\mathbf s}_a\cdot {\mathbf s}_b}N\right)^2f''\left(\frac{{\mathbf s}_a\cdot {\mathbf s}_b}N\right)
        \right]
      \right\}
  \end{aligned}
\end{equation}
We introduce new matrix fields
\begin{align} \label{eq:fields}
  C_{ab}=\frac1N\mathbf s_a\cdot\mathbf s_b &&
  R_{ab}=-i\frac1N\hat{\mathbf s}_a\cdot{\mathbf s}_b &&
  D_{ab}=\frac1N\hat{\mathbf s}_a\cdot\hat{\mathbf s}_b
\end{align}
Their physical meaning is explained in \S\ref{sec:interpretation}.  By
substituting these parameters into the expressions above and then making a
change of variables in the integration from $\mathbf s_a$ and $\hat{\mathbf
s}_a$ to these three matrices, we arrive at the form for the complexity
\begin{equation}
  \begin{aligned}
    \Sigma(E,\mu^*)
    &=\mathcal D(\mu^*)+\hat\beta E-\frac12\hat\mu+
    \lim_{n\to0}\frac1n\left(
      \frac12\hat\mu\operatorname{Tr}C-\mu^*\operatorname{Tr}R\right.\\
    &\hspace{2em}\left.+\frac12\sum_{ab}\left[
        \hat\beta^2f(C_{ab})+(2\hat\beta R_{ab}-D_{ab})f'(C_{ab})
        +R_{ab}^2f''(C_{ab})
      \right]
    +\frac12\ln\det\begin{bmatrix}C&iR\\iR&D\end{bmatrix}
    \right)
  \end{aligned}
\end{equation}
where $\hat\mu$, $\hat\beta$, $C$, $R$ and $D$ must be evaluated at the extrema
of this expression which minimize the complexity. Note that one cannot
\emph{minimize} the complexity with respect to these parameters: there is no
pure variational problem here. Extremizing with respect to $\hat\mu$ is not
difficult, and results in setting the diagonal of $C$ to one, fixing the
spherical constraint. Maintaining $\hat\mu$ in the complexity is useful for
writing down the extremal conditions, but when convenient we will drop the
dependence.

The same information is contained but better expressed in the Legendre
transform
\begin{equation}
  \begin{aligned}
    &G(\hat \beta,\mu^*)
    =\mathcal D(\mu^*)+\\
    &\lim_{n\to0}\frac1n\left(
      -\mu^*\operatorname{Tr}R
      +\frac12\sum_{ab}\left[
        \hat\beta^2f(C_{ab})+(2\hat\beta R_{ab}-D_{ab})f'(C_{ab})
        +R_{ab}^2f''(C_{ab})
      \right]
    +\frac12\ln\det\begin{bmatrix}C&iR\\iR&D\end{bmatrix}
    \right)
  \end{aligned}
\end{equation}
Denoting $r_d \equiv \frac 1 n {\mbox Tr} R$, we can write down the double Legendre transform $K(\hat \beta, r_d)$:
\begin{equation}
  e^{N K(\hat \beta, r_d)} =\int\,dE\,d\mu^* e^{N\left\{\Sigma(E,\mu^*) -\hat\beta E+r_d\mu^* -\mathcal D(\mu^*)\right\}}
\end{equation}
given by
\begin{equation}
  \begin{aligned}
    &K(\hat \beta,r_d)
   = \lim_{n\to0}\frac1n\left(
      \frac12\sum_{ab}\left[
        \hat\beta^2f(C_{ab})+(2\hat\beta R_{ab}-D_{ab})f'(C_{ab})
        +R_{ab}^2f''(C_{ab})
      \right]
    +\frac12\ln\det\begin{bmatrix}C&iR\\iR&D\end{bmatrix}
    \right)
  \end{aligned}
\end{equation}
where the diagonal of $C$ is fixed to one and the diagonal of $R$ is fixed to
$r_d$.  The variable $r_d$ is conjugate to $\mu^*$ and through it to the index
density, while $\hat \beta$ plays the role of an inverse temperature conjugate
to the complexity, that has been used since the beginning of the spin-glass
field. In this way $K(\hat \beta,r_d)$ contains all the information about
saddle densities.

\section{Replica ansatz}
\label{sec:ansatz}

Based on previous work on the Sherrington--Kirkpatrick model and the
equilibrium solution of the spherical model, we expect $C$, and $R$ and $D$ to
be hierarchical matrices in Parisi's scheme. This assumption immediately simplifies the
extremal conditions, since hierarchical matrices commute and are closed under
matrix products and Hadamard products. In particular, the determinant of the block matrix can be written as a determinant of a product,
\begin{equation}
  \ln\det\begin{bmatrix}C&iR\\iR&D\end{bmatrix}=\ln\det(CD+R^2)
\end{equation}
This is straightforward (if strenous) to write down at $k$RSB, since the
product and sum of the hierarchical matrices is still a hierarchical matrix.
The algebra of hierarchical matrices is reviewed in \S\ref{sec:dict}. Using the
product formula \eqref{eq:replica.prod}, one can write down the hierarchical
matrix $CD+R^2$, and then compute the $\ln\det$ using the formula
\eqref{eq:replica.logdet}.

The extremal conditions are given by differentiating the complexity with
respect to its parameters, yielding
\begin{align}
  0&=\frac{\partial\Sigma}{\partial\hat\mu}
  =\frac12(c_d-1) \\
  0&=\frac{\partial\Sigma}{\partial\hat\beta}
  =E+\lim_{n\to0}\frac1n\sum_{ab}\left[\hat\beta f(C_{ab})+R_{ab}f'(C_{ab})\right] \label{eq:cond.b} \\
  0&=\frac{\partial\Sigma}{\partial C}
  =\frac12\left[
    \hat\mu I+\hat\beta^2f'(C)+(2\hat\beta R-D)\odot f''(C)+R\odot R\odot f'''(C)
    +(CD+R^2)^{-1}D
  \right] \label{eq:cond.q} \\
  0&=\frac{\partial\Sigma}{\partial R}
  =-\mu^* I+\hat\beta f'(C)+R\odot f''(C)
  +(CD+R^2)^{-1}R \label{eq:cond.r} \\
  0&=\frac{\partial\Sigma}{\partial D}
  =-\frac12f'(C)
    +\frac12(CD+R^2)^{-1}C \label{eq:cond.d}
\end{align}
where $\odot$ denotes the Hadamard product, or the componentwise product. Equation \eqref{eq:cond.d} implies that
\begin{equation} \label{eq:D.solution}
  D=f'(C)^{-1}-RC^{-1}R
\end{equation}
To these conditions must be added the addition condition that $\Sigma$ is extremal with respect to $x_1,\ldots, x_k$. There is no better way to enforce this condition than to directly differentiate $\Sigma$ with respect to the $x$s, and we have
\begin{equation} \label{eq:cond.x}
  0=\frac{\partial\Sigma}{\partial x_i}\qquad 1\leq i\leq k
\end{equation}
The stationary conditions for the $x$s are the most numerically taxing.

In addition to these equations, we often want to maximize the complexity as a
function of $\mu^*$, to find the most common type of stationary points. These
are given by the condition
\begin{equation} \label{eq:cond.mu}
  0=\frac{\partial\Sigma}{\partial\mu^*}
  =\mathcal D'(\mu^*)-r_d
\end{equation}
Since $\mathcal D(\mu^*)$ is effectively a piecewise function, with different
forms for $\mu^*$ greater or less than $\mu_m$, there are two regimes. When
$\mu^*>\mu_m$ and the critical points are minima, \eqref{eq:cond.mu} implies
\begin{equation} \label{eq:mu.minima}
  \mu^*=\frac1{r_d}+r_df''(1)
\end{equation}
When $\mu^*<\mu_m$ and the critical points are saddles, it implies
\begin{equation} \label{eq:mu.saddles}
  \mu^*=2f''(1)r_d
\end{equation}

It is often useful to have the extremal conditions in a form without matrix
inverses, so that the saddle conditions can be expressed using products
alone. By simple manipulations, the matrix equations
can be written as
\begin{align}
  0&=\left[\hat\beta^2f'(C)+(2\hat\beta R-D)\odot f''(C)+R\odot R\odot f'''(C)+\hat\mu I\right]C+f'(C)D  \\
  0&=\left[\hat\beta f'(C)+R\odot f''(C)-\mu^*I\right]C+f'(C)R \\
  0&=C-f'(C)(CD+R^2)
\end{align}
The right-hand side of each of these equations is also a hierarchical matrix,
since products, Hadamard products, and sums of hierarchical matrices are such.

\section{Supersymmetric solution}
\label{sec:supersymmetric}

The Kac--Rice problem has an approximate supersymmetry, which is found when the
absolute value of the determinant is neglected and the trace of the Hessian is
not fixed. This supersymmetry has been studied in great detail in the
complexity of the Thouless--Anderson--Palmer (TAP) free energy
\cite{Annibale_2003_The, ,Annibale_2003_Supersymmetric,
Annibale_2004_Coexistence, Cavagna_2005_Cavity, Giardina_2005_Supersymmetry}. When the absolute value is dropped, the determinant in \eqref{eq:kac-rice} can be
represented by an integral over Grassmann variables, which yields a complexity
depending on `bosons' and `fermions' that share the supersymmetry. The Ward
identities associated with the supersymmetry imply that $D=\hat\beta R$
\cite{Annibale_2003_The}. Under which conditions can this relationship be
expected to hold? We find that their applicability is limited to a specific
line in the energy and stability plane.

The identity $D=\hat\beta R$ heavily constrains the form that the rest of the
solution can take. Assuming the supersymmetry holds, \eqref{eq:cond.q} implies
\begin{equation}
  0=\hat\mu I+\hat\beta^2f'(C)+\hat\beta R\odot f''(C)+R\odot R\odot f'''(C)+\hat\beta(CD+R^2)^{-1}R
\end{equation}
Substituting \eqref{eq:cond.r} for the factor $(CD+R^2)^{-1}R$, we find substantial cancellation, and finally
\begin{equation} \label{eq:R.diagonal}
  0=(\hat\mu+\mu^*)I+R\odot R\odot f'''(C)
\end{equation}
If $C$ has a nontrivial off-diagonal structure and supersymmetry holds, then
the off-diagonal of $R$ must vanish, and therefore $R=r_dI$. Therefore, a
supersymmetric ansatz is equivalent to a \emph{diagonal} ansatz for both $R$ and $D$.

Supersymmetry has further implications.
Equations \eqref{eq:cond.r} and \eqref{eq:cond.d} can be combined to find
\begin{equation}
  I=R\left[\mu^* I-R\odot f''(C)\right]+(D-\hat\beta R)f'(C)
\end{equation}
Assuming the supersymmetry holds implies that
\begin{equation}
  I=R\left[\mu^* I-R\odot f''(C)\right]
\end{equation}
Understanding that $R$ is diagonal, we find
\begin{equation}
  \mu^*=\frac1{r_d}+r_df''(1)
\end{equation}
which is precisely the condition \eqref{eq:mu.minima} for dominant minima. Therefore, \emph{the
supersymmetric solution counts the most common minima}
\cite{Annibale_2004_Coexistence}. When minima are not the most common type of
stationary point, the supersymmetric solution correctly counts minima that
satisfy \eqref{eq:mu.minima}, but these do not have any other special significance.

Inserting the supersymmetric ansatz $D=\hat\beta R$ and $R=r_dI$, one gets for the complexity
\begin{equation} \label{eq:diagonal.action}
  \begin{aligned}
    \Sigma(E,\mu^*)
    =\mathcal D(\mu^*)
    +
      \hat\beta E-\mu^* r_d
      +\frac12\hat\beta r_df'(1)+\frac12r_d^2f''(1)+\frac12\ln r_d^2
      \\
      +\frac12\lim_{n\to0}\frac1n\left(\hat\beta^2\sum_{ab}f(C_{ab})+\ln\det((\hat\beta/r_d)C+I)\right)
  \end{aligned}
\end{equation}
From here, it is straightforward to see that the complexity vanishes at the
ground state energy. First, in the ground state minima will dominate (even if
they are marginal), so we may assume \eqref{eq:mu.minima}. Then, taking
$\Sigma(E_0,\mu^*)=0$, gives
\begin{equation}
  \hat\beta E_0
  =-\frac12r_d\hat\beta f'(1)-\frac12\lim_{n\to0}\frac1n\left(
      \hat\beta^2\sum_{ab}^nf(C_{ab})
      +\ln\det(\hat\beta r_d^{-1} C+I)
  \right)
\end{equation}
which is precisely the ground state energy predicted by the equilibrium
solution \eqref{eq:ground.state.free.energy} with $r_d=z$,
$\hat\beta=\tilde\beta$, and $C=\tilde Q$.

{\em Therefore a $(k-1)$RSB ansatz in
Kac--Rice will predict the correct ground state energy for a model whose
equilibrium state at small temperatures is $k$RSB } Moreover, there is an
exact correspondence between the saddle parameters of each.  If the equilibrium
is given by a Parisi matrix with parameters $x_1,\ldots,x_k$ and
$q_1,\ldots,q_k$, then the parameters $\hat\beta$, $r_d$, $d_d$, $\tilde
x_1,\ldots,\tilde x_{k-1}$, and $c_1,\ldots,c_{k-1}$ for the
complexity in the ground state are
\begin{align}\label{eq:equilibrium.complexity.map}
  \hat\beta=\lim_{\beta\to\infty}\beta x_k
  &&
  \tilde x_i=\lim_{\beta\to\infty}\frac{x_i}{x_k}
  &&
  c_i=\lim_{\beta\to\infty}q_i
  &&
  r_d=\lim_{\beta\to\infty}\beta(1-q_k)
  &&
  d_d=\hat\beta r_d
\end{align}
Unlike the case for the TAP complexity, this correspondence between landscape complexity
and equilibrium solutions only exists at the ground state. We will see in our
examples in \S\ref{sec:examples} that there appears to be little correspondence
between these parameters away from the ground state.

The supersymmetric solution produces the correct complexity for the ground
state and for a class of minima, including dominant ones. Moreover, it produces the correct parameters
for the fields $C$, $R$, and $D$ at those points. This is an important foothold
in the problem of computing the general complexity. The full saddle point
equations at $k$RSB are not very numerically stable, and a `good' saddle point
has a typically small radius of convergence under methods like Newton's
algorithm. With the supersymmetric solution in hand, it is possible to take
small steps in the parameter space to find non-supersymmetric numeric
solutions, each time ensuring the initial conditions for the solver are
sufficiently close to the correct answer. This is the strategy we use in
\S\ref{sec:examples}.

\section{Full replica symmetry breaking}
\label{sec:frsb}

This reasoning applies equally well to FRSB systems. In the end, when the
limit of $n\to0$ is taken, each matrix field can be represented in the
canonical way by its diagonal and a continuous function on the domain $[0,1]$
which parameterizes each of its rows, with
\begin{align}
  C\;\leftrightarrow\;[c_d, c(x)]
  &&
  R\;\leftrightarrow\;[r_d, r(x)]
  &&
  D\;\leftrightarrow\;[d_d, d(x)]
\end{align}
The algebra of hierarchical matrices under this continuous parameterization is
reviewed in \S\ref{sec:dict}.  With these substitutions, the complexity becomes
\begin{equation}
  \begin{aligned}
    \Sigma(E,\mu^*)
    &=\mathcal D(\mu^*)+\hat\beta E-\mu^*r_d +\frac12\left[\hat\beta^2f(1)+(2\hat\beta r_d-d_d)f'(1)+r_d^2f''(1)\right]\\
    &-\frac12\int_0^1dx\left[
      \hat\beta^2f(c(x))+(2\hat\beta r(x)-d(x))f'(c(x))
      +r(x)^2f''(c(x))
      \right]
      +\frac12\lim_{n\to0}\frac1n\ln\det(CD+R^2)
  \end{aligned}
\end{equation}
The formula for the determinant is complicated, and can be found by using the
product formula \eqref{eq:cont.replica.prod} to write $CD$ and $R^2$, summing
them, and finally using the $\ln\det$ formula \eqref{eq:replica.det.cont}.
The saddle point equations take the form
\begin{align}
  0&=\hat\mu c(x)+\Big[\big(\hat\beta^2(f'\circ c)+(2\hat\beta r-d)(f''\circ c)+r^2(f'''\circ c)\big)\ast c\Big](x)+\big((f'\circ c)\ast d\big)(x) \label{eq:extremum.c} \\
  0&=-\mu^* c(x)+\Big[\big(\hat\beta(f'\circ c)+r*(f''\circ c)\big)\ast c\Big](x)+\big((f'\circ c)\ast r\big)(x) \label{eq:extremum.r} \\
  0&=c(x)-\big((f'\circ c)\ast(c\ast d+r\ast r)\big)(x) \label{eq:extremum.d}
\end{align}
where $(ab)(x)=a(x)b(x)$ denotes the hadamard product, $(a\ast b)(x)$ denotes
the functional parameterization of the diagonal of the product of hierarchical
matrices $AB$ defined in \eqref{eq:cont.replica.prod}, and $(a\circ
b)(x)=a(b(x))$ denotes composition.

\subsection{Supersymmetric complexity}

Using standard manipulations, one finds also a continuous version of the
supersymmetric complexity
\begin{equation} \label{eq:functional.action}
    \Sigma(E,\mu^*)
    =\mathcal D(\mu^*)
    +
      \hat\beta E-\mu^* r_d
      +\frac12\left(\hat\beta r_df'(1)+r_d^2f''(1)+\ln r_d^2\right)
      +\frac12\int_0^1dq\,\left(
        \hat\beta^2f''(q)\chi(q)+\frac1{\chi(q)+r_d/\hat\beta}
      \right)
\end{equation}
where $\chi(q)=\int_1^qdq'\int_0^{q'}dq''\,P(q)$ for $P(q)$ the distribution of elements in a row of $C$, as in the equilibrium case. Like in the equilibrium case, $\chi$ must be concave, monotonically decreasing, and have $\chi(1)=0$, $\chi'(1)=-1$.

First, we use this solution to inspect the ground state of a full RSB system.
We know from the equilibrium that in the ground state $\chi$ is continuous in
the whole range of $q$. Therefore, the saddle solution found by extremizing
\begin{equation}
  0=\frac{\delta\Sigma}{\delta\chi(q)}=\frac12\hat\beta^2f''(q)-\frac12\frac1{(\chi(q)+r_d/\hat\beta)^2}
\end{equation}
over all functions $\chi$. This gives
\begin{equation}
  \chi_0(q\mid\hat\beta,r_d)=\frac1{\hat\beta}\left(f''(q)^{-1/2}-r_d\right)
\end{equation}
Satisfying the boundary conditions requires $r_d=f''(1)^{-1/2}$ and $\hat\beta=\frac12f'''(1)/f''(1)^{3/2}$.
This in turn implies $\mu^*=\frac1{r_d}+f''(1)r_d=\sqrt{4f''(1)}=\mu_m$.
Therefore, the FRSB ground state is always marginal, as excepted. It is
straightforward to check that these conditions are indeed a saddle of the
complexity.
This has several implications. First, other than the ground state, there are
\emph{no} energies at which minima are most numerous; saddles always dominate.
As we will see, stable minima are numerous at energies above the ground state,
but these vanish at the ground state.

Away from the ground state, this expression still correctly counts a class of
non-dominant minima. However, like in the equilibrium solution, the function
$\chi$ which produces an extremal value is not smooth in the entire range
$[0,1]$, but adopts a piecewise form
\begin{equation}
  \chi(q)=\begin{cases}
    \chi_0(q\mid\hat\beta,r_d) & q\leq q_\textrm{max} \\
    1-q & \text{otherwise}
  \end{cases}
\end{equation}
With this ansatz, the complexity must be extremized with respect to $r_d$ and
$\hat\beta$, while simultaneously ensuring that $q_\textrm{max}$ is such that
$\chi(q)$ is continuous, that is, that
$\chi_0(q_\textrm{max}\mid\hat\beta,r_d)=1-q_\textrm{max}$. The significance of
the minima counted by this method is unclear, but they do represent a nodal
line in the off-diagonal parts of $R$ and $D$. Since, as usual, $\chi(q)$ is
related to $c(x)$ by $-\chi'(c(x))=x$, there is a corresponding $x_\textrm{max}$ given by
\begin{equation}
  x_\textrm{max}=-\chi'(q_\textrm{max})=\frac1{2\hat\beta}\frac{f'''(q_\textrm{max})}{f''(q_\textrm{max})^{3/2}}
\end{equation}

\subsection{Expansion near the transition}
\label{subsec:expansion}

Working with the continuum equations away from the
supersymmetric solution is not generally tractable. However, there is another
point where they can be treated analytically: near the onset of replica
symmetry breaking. Here, the off-diagonal components of $C$, $R$, and $D$ are
expected to be small. In particular, we expect the functions $c(x)$, $r(x)$, and $d(x)$
to approach zero at the transition, and moreover take the piecewise linear form
\begin{align}
  c(x)=\begin{cases}\bar cx&x\leq x_\mathrm{max}\\\bar cx_\mathrm{max}&\text{otherwise}\end{cases}
                           &&
  r(x)=\begin{cases}\bar rx&x\leq x_\mathrm{max}\\\bar rx_\mathrm{max}&\text{otherwise}\end{cases}
                           &&
  d(x)=\begin{cases}\bar dx&x\leq x_\mathrm{max}\\\bar dx_\mathrm{max}&\text{otherwise}\end{cases}
\end{align}
with $x_\mathrm{max}$ vanishing at the transition, with the slopes $\bar c$,
$\bar r$, and $\bar d$ remaining nonzero. This ansatz is informed both by the
experience of the equilibrium solution, and by empirical observation within the
numerics of \S\ref{sec:examples}

Given this ansatz, we take the equations \eqref{eq:extremum.c},
\eqref{eq:extremum.r}, and \eqref{eq:extremum.d}, which are true for any $x$,
and integrate them over $x$. We then expand the result about small
$x_\mathrm{max}$ to linear order in $x_\mathrm{max}$. Equation
\eqref{eq:extremum.r} depends linearly on $\bar r$ to all orders, and therefore
$\bar r$ can be found in terms of $\bar c$, yielding
\begin{equation}
  \begin{aligned}
    \frac{\bar r}{\bar c}
    &=
    -\hat\beta-\frac1{f'(1)+f''(0)}\left(r_d(f''(0)+f''(1))-\mu^*\right)+O(x_\mathrm{max})
  \end{aligned}
\end{equation}
Likewise, \eqref{eq:extremum.d} depends linearly on $\bar d$ to all orders, and can be solved to give
\begin{equation}
  \begin{aligned}
    \frac{\bar d}{\bar c}
    &=-2r_d\frac{\bar r}{\bar c}-\frac1{f'(1)}(r_d^2f''(0)+d_d(f'(1)+f''(0))-1)+O(x_\mathrm{max})
  \end{aligned}
\end{equation}
The equations cannot be used to find the value of $\bar c$ without going to
higher order in $x_\mathrm{max}$, but the transition line can be determined by
examining the stability of the replica symmetric complexity. First, we expand
the full form for the complexity about small $x_\textrm{max}$ in the same way
as we expand the extremal conditions, using \eqref{eq:replica.det.cont} to
treat the determinant. To quadratic order, this gives
\begin{equation}
  \begin{aligned}
  &\Sigma(E,\mu^*)
  =\mathcal D(\mu^*)+\hat\beta E-\mu r_d+\frac12\left[\hat\beta^2f(1)+(2\hat\beta r_d-d_d)f'(1)+r_d^2f''(1)\right]+\frac12\ln(d_d+r_d^2) \\
  &-\frac12\left[
    \frac12\hat\beta^2\bar c^2f''(0)+(2\hat\beta\bar r-\bar d)\bar cf''(0)+\bar r^2f''(0)
    -\frac{\bar d^2-2d_d\bar r^2+d_d^2\bar c^2+4r_d\bar r(\bar d+d_d\bar c)-2r_d^2(\bar c\bar d+\bar r^2)}{2(d_d+r_d^2)^2}
  \right]x_\textrm{max}^2
  \end{aligned}
\end{equation}
The spectrum of the Hessian of $\Sigma$ with evaluated at the RS
solution gives its stability with respect to these functional perturbations. When the
values of $\bar r$ and $\bar d$ above are substituted into the Hessian and
$\hat\beta$, $r_d$, and $d_d$ are evaluated at their RS values, the eigenvalue
of interest takes the form
\begin{equation}
  \lambda
  =-\bar c^2\frac{(f'(1)-2f(1))^2(f'(1)-f''(0))f''(0)}{2(f'(1)+f''(0))(f'(1)^2-f(1)(f'(1)+f''(1)))^2}(\mu^*-\mu^*_+(E))(\mu^*-\mu^*_-(E))
\end{equation}
where
\begin{equation} \label{eq:mu.transition}
  \mu^*_\pm(E)
  =\pm\frac{(f'(1)+f''(0))(f'(1)^2-f(1)(f'(1)+f''(1)))}{(2f(1)-f'(1))f'(1)f''(0)^{-1/2}}
  -\frac{f''(1)-f'(1)}{f'(1)-2f(1)}E
\end{equation}
This eigenvalue changes sign when $\mu^*$ crosses $\mu^*_\pm(E)$.  We expect
that this is the line of stability for the replica symmetric solution when the
transition is RS-FRSB. The numerics in \S\ref{sec:examples} bear this out.

\section{General solution: examples}
\label{sec:examples}

Though we have only written down an easily computable complexity along a
specific (and often uninteresting) line in energy and stability, this
computable (supersymmetric) solution gives a numeric foothold for computing the
complexity in the rest of that space. First,
\eqref{eq:ground.state.free.energy.cont} is \emph{maximized} with respect
to its parameters, since the equilibrium solution is equivalent to a
variational problem. Second, the mapping \eqref{eq:equilibrium.complexity.map}
is used to find the corresponding Kac--Rice saddle parameters in the ground
state. With these parameters in hand, small steps are then made in energy $E$
or stability $\mu$, after which known these values are used as the initial condition
for a saddle-finding problem. In this section, we use this basic numeric idea
to map out the complexity for two representative examples: a model with a
2RSB equilibrium ground state and therefore 1RSB complexity in its
vicinity, and a model with a FRSB equilibrium ground state, and therefore
FRSB complexity as well.

\subsection{1RSB complexity}

It is known that by choosing a covariance $f$ as the sum of polynomials with
well-separated powers, one develops 2RSB in equilibrium. This should correspond
to 1RSB in Kac--Rice. For this example, we take
\begin{equation}
  f(q)=\frac12\left(q^3+\frac1{16}q^{16}\right)
\end{equation}
established to have a 2RSB ground state \cite{Crisanti_2011_Statistical}.
With this covariance, the model sees a replica symmetric to 1RSB transition at
$\beta_1=1.70615\ldots$ and a 1RSB to 2RSB transition at
$\beta_2=6.02198\ldots$. At these transitions, the average energies in equilibrium are
$\langle E\rangle_1=-0.906391\ldots$ and $\langle E\rangle_2=-1.19553\ldots$,
respectively, and the ground state energy is $E_0=-1.287\,605\,530\ldots$.
Besides these typical equilibrium energies, an energy of special interest for
looking at the landscape topology is the \emph{algorithmic threshold}
$E_\mathrm{alg}$, defined by the lowest energy reached by local algorithms like
approximate message passing \cite{ElAlaoui_2020_Algorithmic,
ElAlaoui_2021_Optimization}. In the spherical models, this has been proven to
be
\begin{equation}
  E_{\mathrm{alg}}=-\int_0^1dq\,\sqrt{f''(q)}
\end{equation}
For full RSB systems, $E_\mathrm{alg}=E_0$ and the algorithm can reach the
ground state energy. For the pure $p$-spin models,
$E_\mathrm{alg}=E_\mathrm{th}$, where $E_\mathrm{th}$ is the energy at which
marginal minima are the most common stationary points. Something about the
topology of the energy function might be relevant to where this algorithmic threshold
lies. For the $3+16$ model at hand, $E_\mathrm{alg}=-1.275\,140\,128\ldots$.

In this model, the RS complexity gives an inconsistent answer for the
complexity of the ground state, predicting that the complexity of minima
vanishes at a higher energy than the complexity of saddles, with both at a
lower energy than the equilibrium ground state. The 1RSB complexity resolves
these problems, predicting the same ground state as equilibrium and with a ground state stability $\mu_0=6.480\,764\ldots>\mu_m$. It predicts that the
complexity of marginal minima (and therefore all saddles) vanishes at
$E_m=-1.287\,605\,527\ldots$, which is very slightly greater than $E_0$. Saddles
become dominant over minima at a higher energy $E_\mathrm{th}=-1.287\,575\,114\ldots$.
The 1RSB complexity transitions to a RS description for dominant stationary
points at an energy $E_1=-1.273\,886\,852\ldots$. The highest energy for which
the 1RSB description exists is $E_\mathrm{max}=-0.886\,029\,051\ldots$

\begin{figure}
  \includegraphics[width=\textwidth]{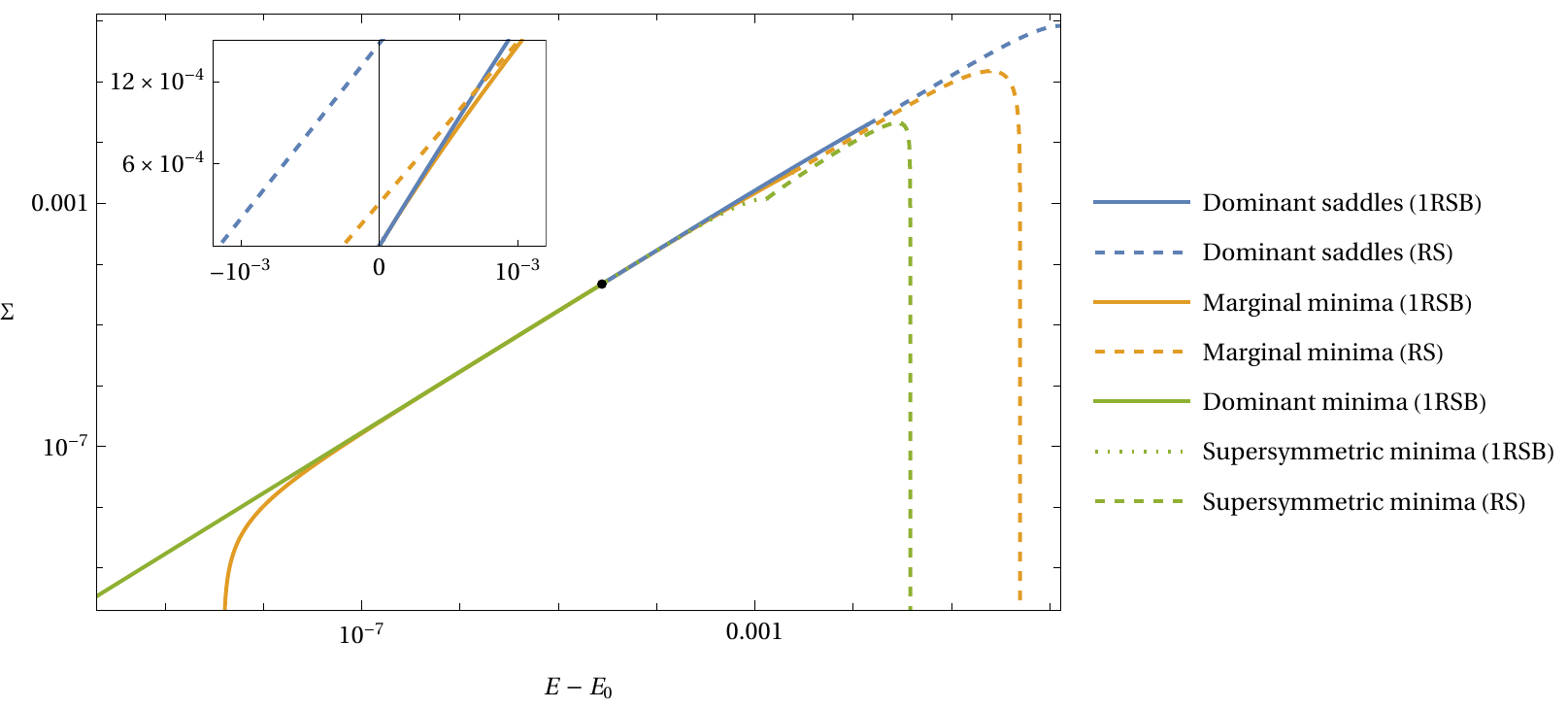}

  \caption{
    Complexity of dominant saddles (blue), marginal minima (yellow), and
    dominant minima (green) of the $3+16$ model. Solid lines show the result of
    the 1RSB ansatz, while the dashed lines show that of a RS ansatz. The
    complexity of marginal minima is always below that of dominant critical
    points except at the black dot, where they are dominant.
    The inset shows a region around the ground state and the fate of the RS solution.
  } \label{fig:2rsb.complexity}
\end{figure}

\begin{figure}
  \centering
  \hspace{-1em}
  \includegraphics{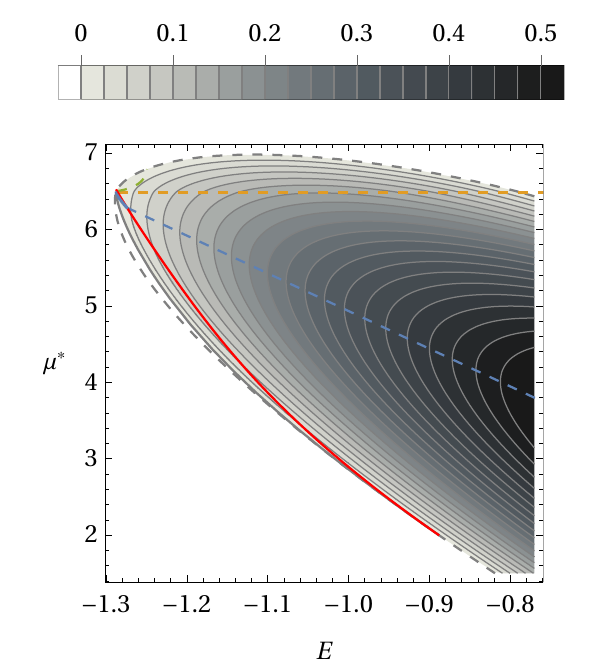}
  \hspace{-1em}
  \includegraphics{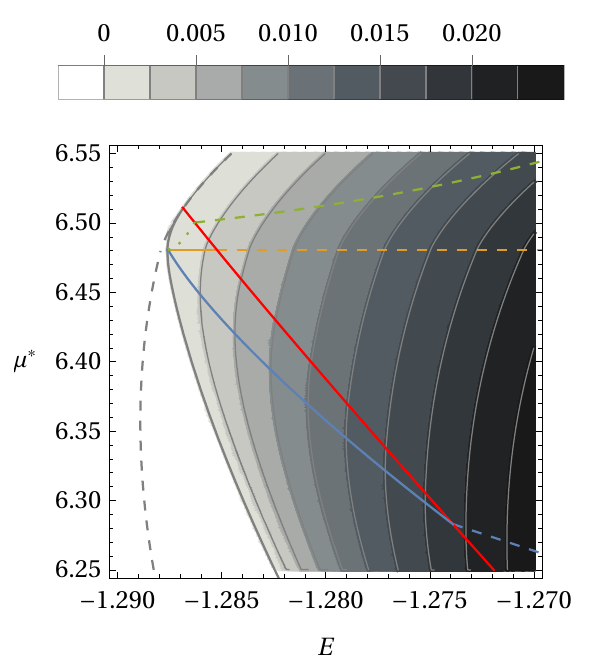}
  \raisebox{3em}{\includegraphics[width=0.27\textwidth]{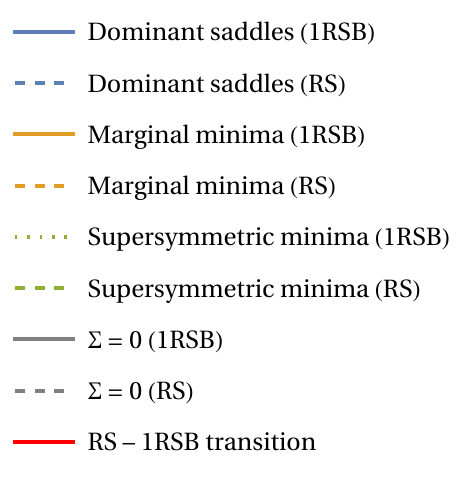}}

  \caption{
    Complexity of the $3+16$ model in the energy $E$ and stability $\mu^*$
    plane. The right shows a detail of the left. Below the yellow marginal line
    the complexity counts saddles of increasing index as $\mu^*$ decreases.
    Above the yellow marginal line the complexity counts minima of increasing
    stability as $\mu^*$ increases.
  } \label{fig:2rsb.contour}
\end{figure}

The complexity as a function of energy difference from the ground state is
plotted in Fig.~\ref{fig:2rsb.complexity}. In that figure, the complexity is
plotted for dominant minima and saddles, marginal minima, and supersymmetric
minima. A contour plot of the complexity as a function of energy $E$ and
stability $\mu$ is shown in Fig.~\ref{fig:2rsb.contour}. That plot also shows
the RS--1RSB transition line in the complexity. For minima, the complexity does
not inherit a 1RSB description until the energy is with in a close vicinity of
the ground state. On the other hand, for high-index saddles the complexity
becomes described by 1RSB at quite high energies. This suggests that when
sampling a landscape at high energies, high index saddles may show a sign of
replica symmetry breaking when minima or inherent states do not.

\begin{figure}
  \centering
  \includegraphics[width=\textwidth]{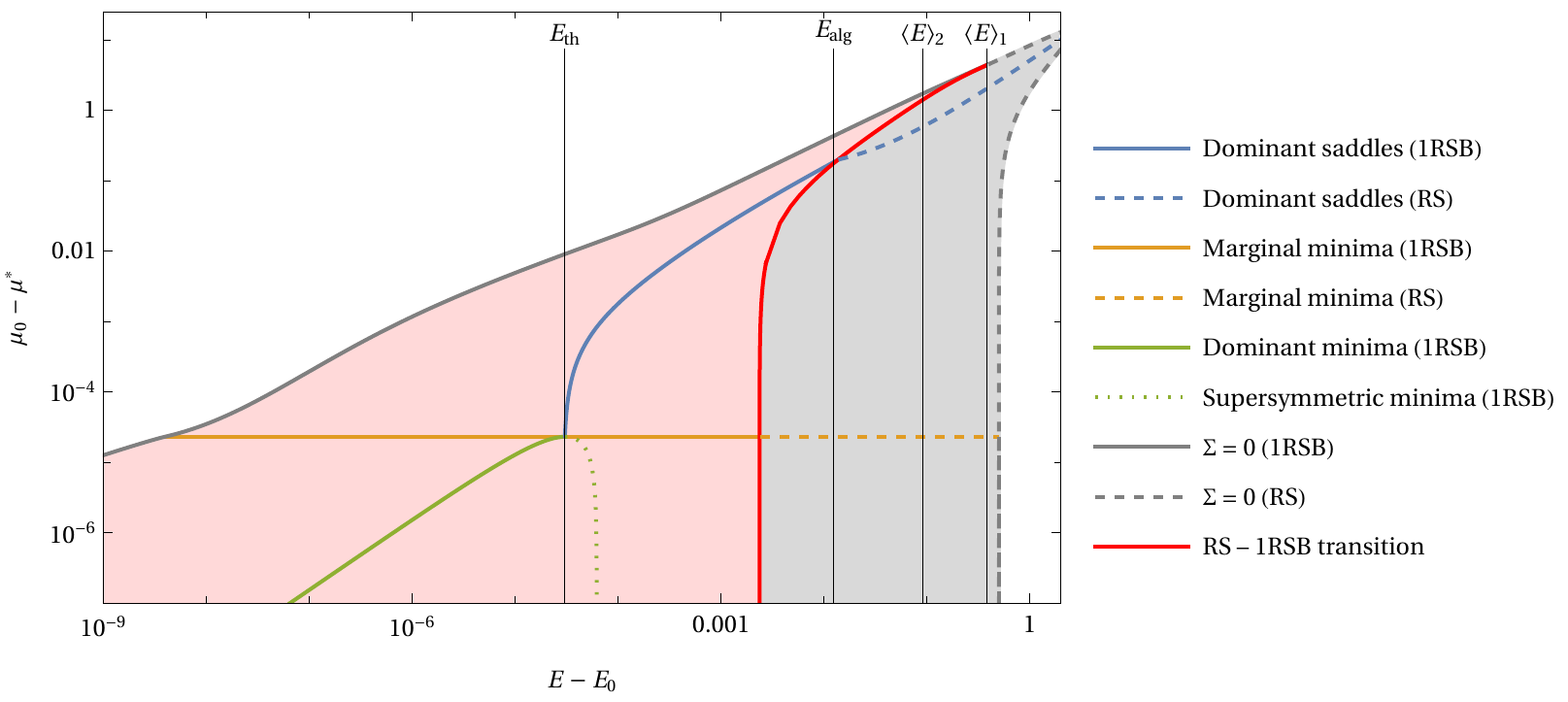}

  \caption{
    Detail of the `phases' of the $3+16$ model complexity as a function of
    energy and stability. Above the yellow marginal stability line the
    complexity counts saddles of fixed index, while below that line it counts
    minima of fixed stability. The shaded red region shows places where the
    complexity is described by the 1RSB solution, while the shaded gray region
    shows places where the complexity is described by the RS solution. In white
    regions the complexity is zero. Several interesting energies are marked
    with vertical black lines: the traditional `threshold' $E_\mathrm{th}$
    where minima become most numerous, the algorithmic threshold
    $E_\mathrm{alg}$ that bounds the performance of smooth algorithms, and the
    average energies at the $2$RSB and $1$RSB equilibrium transitions $\langle
    E\rangle_2$ and $\langle E\rangle_1$, respectively. Though the figure is
    suggestive, $E_\mathrm{alg}$ lies at slightly lower energy than the termination of the RS
    -- 1RSB transition line.
  } \label{fig:2rsb.phases}
\end{figure}

Fig.~\ref{fig:2rsb.phases} shows a different detail of the complexity in the
vicinity of the ground state, now as functions of the energy difference and
stability difference from the ground state. Several of the landmark energies
described above are plotted, alongside the boundaries between the `phases.'
Though $E_\mathrm{alg}$ looks quite close to the energy at which dominant
saddles transition from 1RSB to RS, they differ by roughly $10^{-3}$, as
evidenced by the numbers cited above. Likewise, though $\langle E\rangle_1$
looks very close to $E_\mathrm{max}$, where the 1RSB transition line
terminates, they too differ. The fact that $E_\mathrm{alg}$ is very slightly
below the place where most saddle transition to 1RSB is suggestive; we
speculate that an analysis of the typical minima connected to these saddles by
downward trajectories will coincide with the algorithmic limit. An analysis of
the typical nearby minima or the typical downward trajectories from these
saddles at 1RSB is warranted \cite{Ros_2019_Complex, Ros_2021_Dynamical}. Also
notable is that $E_\mathrm{alg}$ is at a significantly higher energy than
$E_\mathrm{th}$; according to the theory, optimal smooth algorithms in this
model stall in a place where minima are exponentially subdominant.

\begin{figure}
  \centering
  \includegraphics{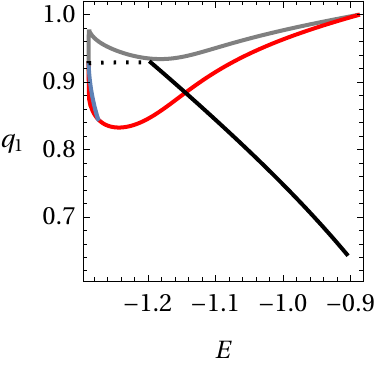}\hfill
  \includegraphics{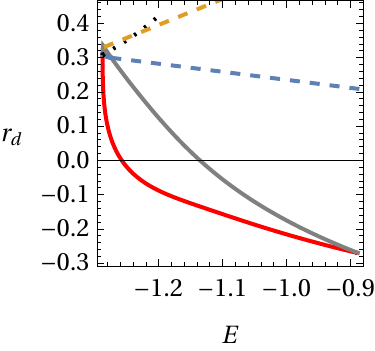}\hfill
  \includegraphics{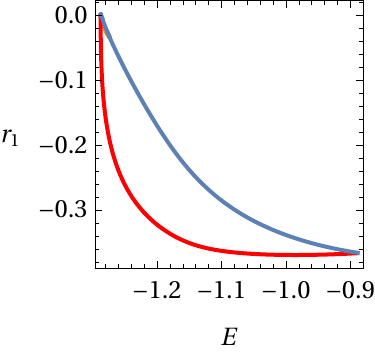}\hfill
  \raisebox{1em}{\includegraphics[width=0.24\textwidth]{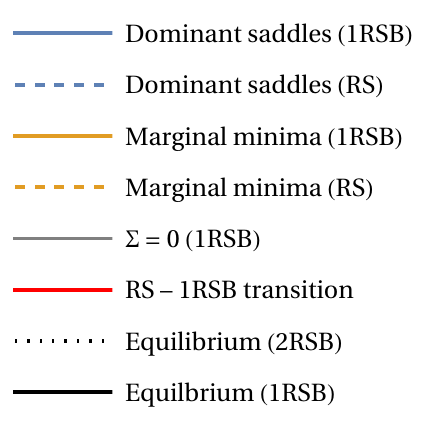}} \vspace{0.5em}\\
  \includegraphics{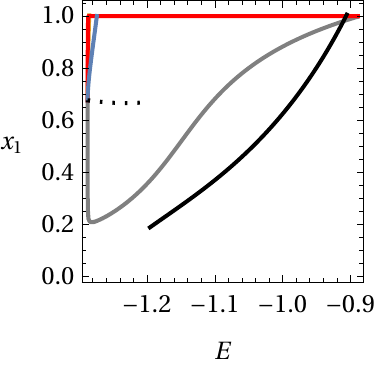}\hfill
  \includegraphics{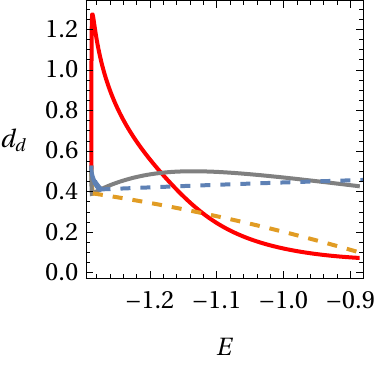}\hfill
  \includegraphics{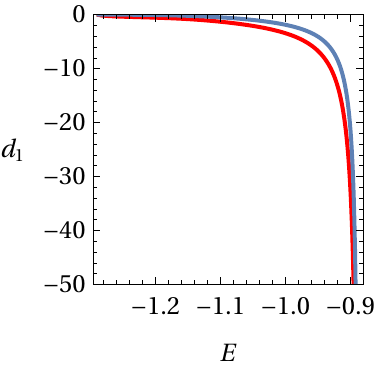}\hfill
  \includegraphics{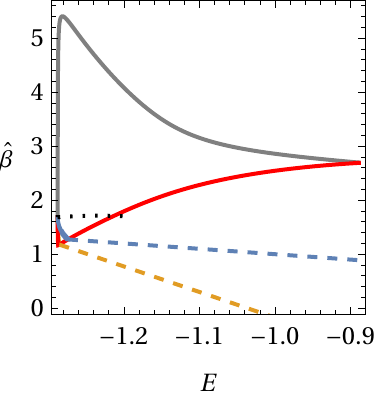}

  \caption{
    Comparison of the saddle point parameters for the $3+16$ model along
    different trajectories in the energy and stability space, and with the
    equilibrium values (when they exist) at the same value of average energy $\langle
    E\rangle$.
  } \label{fig:2rsb.comparison}
\end{figure}

Fig.~\ref{fig:2rsb.comparison} shows the saddle parameters for the $3+16$
system for notable species of stationary points, notably the most common, the
marginal ones, those with zero complexity, and those on the transition line.
When possible, these are compared with the same expressions in the equilibrium
solution at the same average energy. Besides the agreement at the ground state
energy, there seems to be little correlation between the equilibrium and
complexity parameters.

Of specific note is what happens to $d_1$ as the 1RSB phase boundary for the
complexity meets the zero complexity line. Here, $d_1$ diverges like
\begin{equation}
  d_1=-\left(\frac1{f'(1)}-(d_d+r_d^2)\right)(1-x_1)^{-1}+O(1)
\end{equation}
while $x_1$ and $q_1$ both go to one. Note that this is the only place along
the phase boundary where $q_1$ goes to one. The significance of this critical
point in the complexity of high-index saddles in worth further study.

\subsection{Full RSB complexity}

If the covariance $f$ is chosen to be concave, then one develops FRSB in equilibrium. To this purpose, we choose
\begin{equation}
  f(q)=\frac12\left(q^2+\frac1{16}q^4\right)
\end{equation}
also studied before in equilibrium \cite{Crisanti_2004_Spherical, Crisanti_2006_Spherical}. Because the ground state is FRSB, for this model
\begin{equation}
  E_0=E_\mathrm{alg}=E_\mathrm{th}=-\int_0^1dq\,\sqrt{f''(q)}=-1.059\,384\,319\ldots
\end{equation}
In the equilibrium solution, the transition temperature from RS to FRSB is $\beta_\infty=1$, with corresponding average energy $\langle E\rangle_\infty=-0.53125\ldots$.

Along the supersymmetric line, the FRSB solution can be found in full, exact
functional form.  To treat the FRSB away from this line numerically, we resort to
finite $k$RSB approximations.  Since we are not trying to find the actual
$k$RSB solution, but approximate the FRSB one, we drop the extremal condition
\eqref{eq:cond.x} for $x_1,\ldots,x_k$ and instead set
\begin{equation}
  x_i=\left(\frac i{k+1}\right)x_\textrm{max}
\end{equation}
and extremize over $x_\textrm{max}$ alone. This dramatically simplifies the
equations that must be solved to find solutions. In the results that follow, a
20RSB approximation is used to trace the dominant saddles and marginal minima, while
a 5RSB approximation is used to trace the (much longer) boundaries of the
complexity.

\begin{figure}
  \centering
  \includegraphics[width=\textwidth]{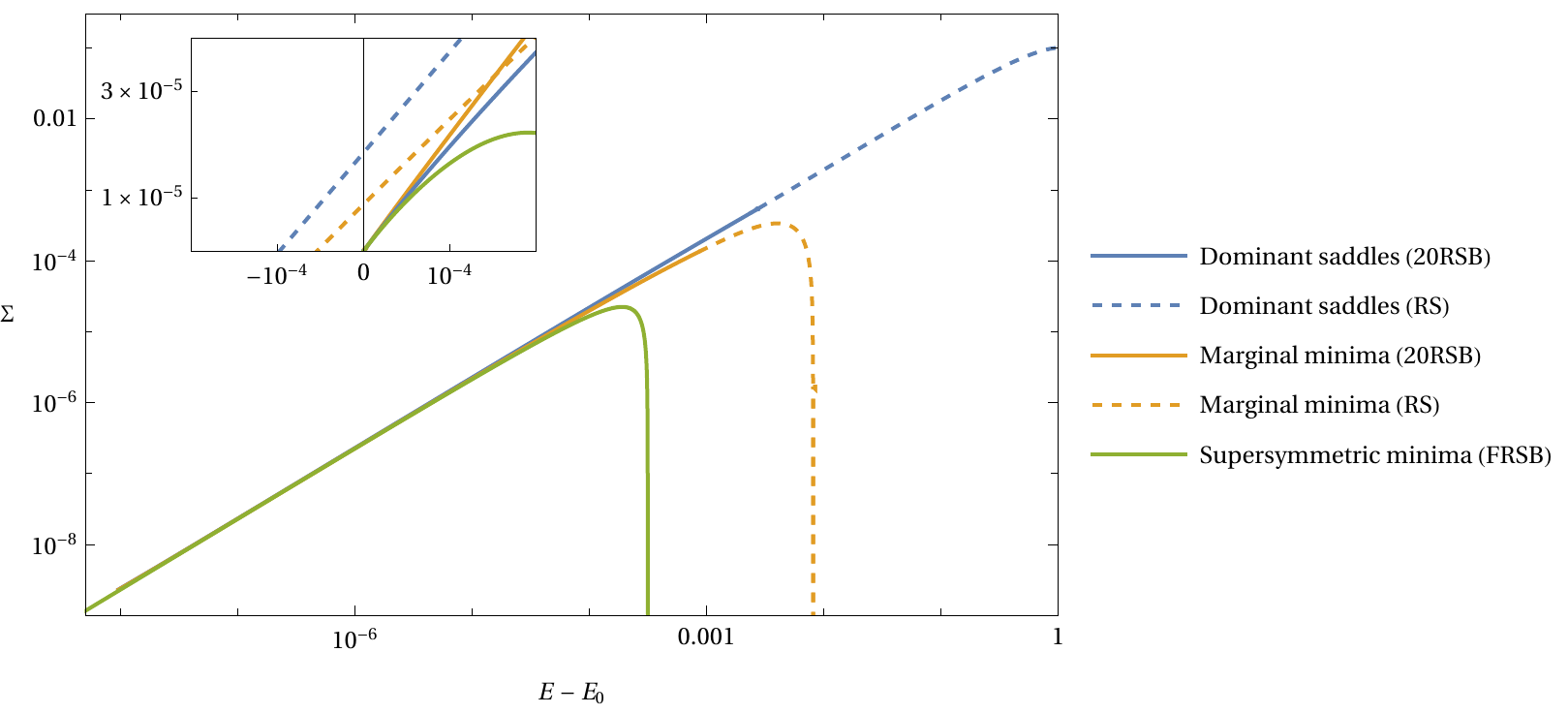}
  \caption{
    The complexity $\Sigma$ of the mixed $2+4$ spin model as a function of
    distance $\Delta E=E-E_0$ of the ground state. The
    solid blue line shows the complexity of dominant saddles given by the FRSB
    ansatz, and the solid yellow line shows the complexity of marginal minima.
    The dashed lines show the same for the annealed complexity. The inset shows
    more detail around the ground state.
  } \label{fig:frsb.complexity}
\end{figure}

\begin{figure}
  \centering
  \includegraphics[width=\textwidth]{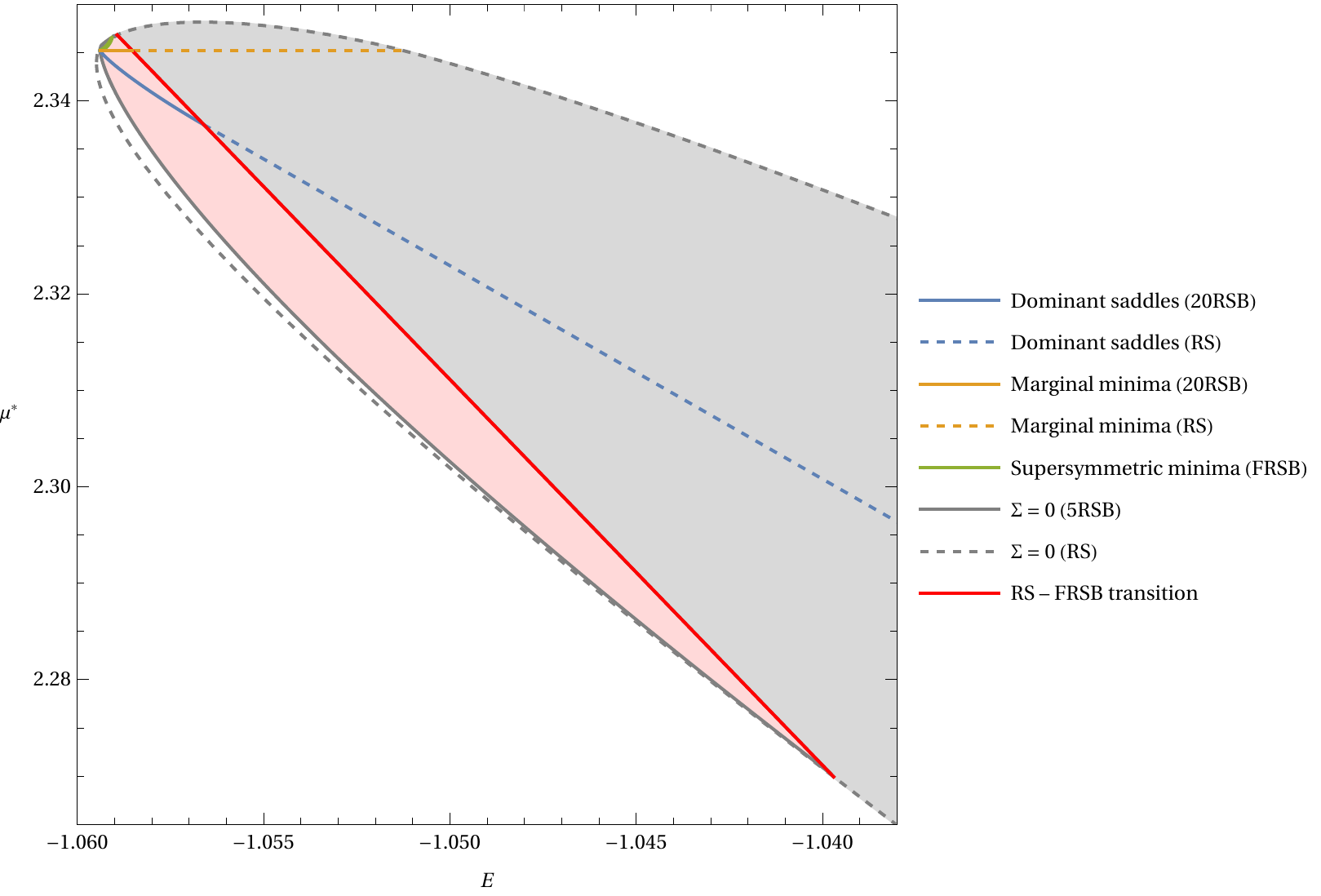}
  \caption{
    `Phases' of the complexity for the $2+4$ model in the energy $E$ and
    stability $\mu^*$ plane. The region shaded gray shows where the RS solution
    is correct, while the region shaded red shows that where the FRSB solution
    is correct. The white region shows where the complexity is zero.
  } \label{fig:frsb.phases}
\end{figure}

Fig.~\ref{fig:frsb.complexity} shows the complexity for this model as a
function of energy difference from the ground state for several notable
trajectories in the energy and stability plane. Fig.~\ref{fig:frsb.phases}
shows these trajectories, along with the phase boundaries of the complexity in
this plane. Notably, the phase boundary predicted by \eqref{eq:mu.transition}
correctly predicts where all of the finite $k$RSB approximations terminate.
Like the 1RSB model in the previous subsection, this phase boundary is oriented
such that very few, low energy, minima are described by a FRSB solution, while
relatively high energy saddles of high index are also. Again, this suggests
that studying the mutual distribution of high-index saddle points might give
insight into lower-energy symmetry breaking in more general contexts.

\begin{figure}
  \raggedright
  \hspace{1em}\includegraphics{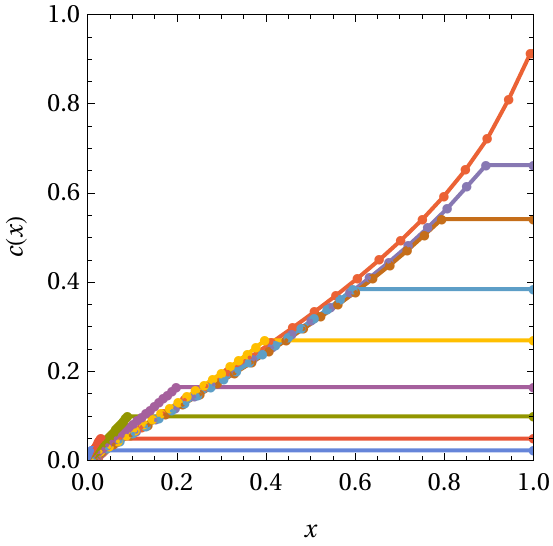}
  \hspace{0.5em}
  \includegraphics{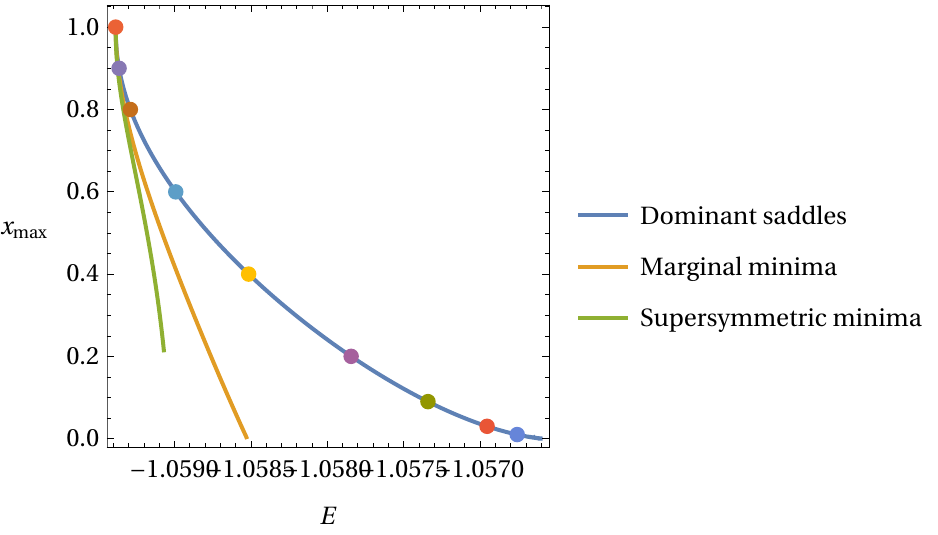}
  \\\vspace{1em}
  \includegraphics{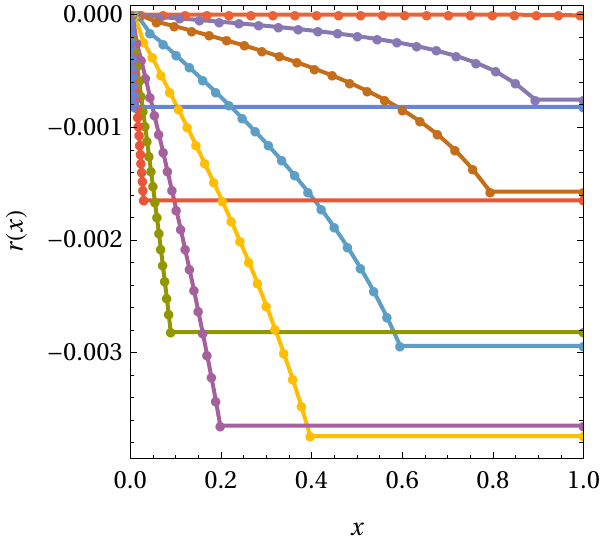}
  \hspace{1em}
  \includegraphics{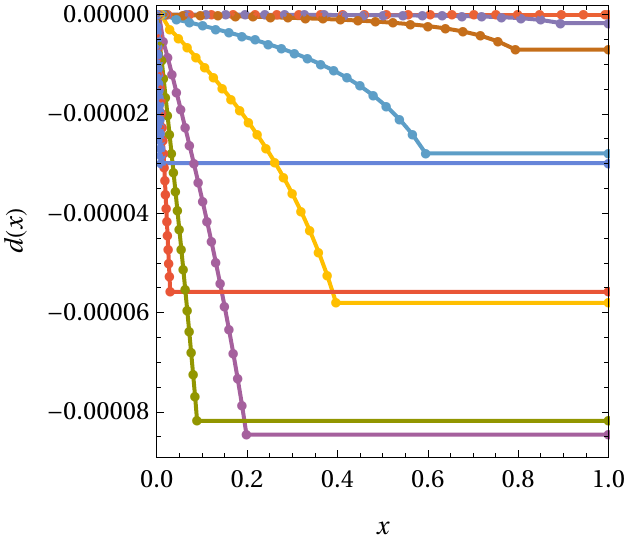}

  \caption{
    $x_\textrm{max}$ as a function of $E$ for several trajectories of interest,
    along with examples of the 20RSB approximations of the functions $c(x)$,
    $r(x)$, and $d(x)$ along the dominant saddles. Colors of the approximate
    functions correspond to the points on the $x_\textrm{max}$ plot. The
    supersymmetric line terminates where the complexity reaches zero, which
    happens inside the FRSB phase.
  } \label{fig:24.func}
\end{figure}

Fig.~\ref{fig:24.func} shows the value of $x_\textrm{max}$ along several
trajectories of interest. Everywhere along the transition line,
$x_\textrm{max}$ continuously goes to zero. Examples of our 20RSB
approximations of the continuous functions $c(x)$, $r(x)$, and $d(x)$ are also
shown. As expected, these functions approach linear ones as $x_\textrm{max}$
goes to zero with finite slopes.

\section{Interpretation}
\label{sec:interpretation}

Let $\langle A\rangle$ be the average of any function $A$ over stationary
points with given $E$ and $\mu^*$, i.e.,
\begin{equation}
  \langle A\rangle
  =\frac1{\mathcal N}\sum_{\mathbf s\in\mathcal S}A(\mathbf s)
  =\frac1{\mathcal N}
  \int d\nu(\mathbf s)\,A(\mathbf s)
\end{equation}
with
\begin{equation}
  d\nu(\mathbf s)=d\mathbf s\,d\mu\,\delta\big(\tfrac12(\|\mathbf s\|^2-N)\big)\,\delta\big(\nabla H(\mathbf s,\mu)\big)\,\big|\det\operatorname{Hess}H(\mathbf s,\mu)\big|
    \delta\big(NE-H(\mathbf s)\big)\delta\big(N\mu^*-\operatorname{Tr}\operatorname{Hess}H(\mathbf s,\mu)\big)
\end{equation}
the Kac--Rice measure. Note that this definition of the angle brackets, which
is in analogy with the typical equilibrium average, is not the same as that
used in \S\ref{subsec:expansion} for averaging over the off-diagonal elements
of a hierarchical matrix. The fields $C$, $R$, and $D$ defined in
\eqref{eq:fields} can be related to certain averages of this type.

\subsection{\textit{C}: distribution of overlaps}

First consider $C$, which has an interpretation nearly identical to that of Parisi's
$Q$ matrix of overlaps in the equilibrium case. Its off-diagonal corresponds to
the probability distribution $P(q)$ of the overlaps $q=(\mathbf s_1\cdot\mathbf
s_2)/N$ between stationary points.  Let $\mathcal S$ be the set of all
stationary points with given energy density and index. Then
\begin{equation}
P(q)\equiv\frac1{\mathcal N^2}\sum_{\mathbf s_1\in\mathcal S}\sum_{\mathbf s_2\in\mathcal S}\delta\left(\frac{\mathbf s_1\cdot\mathbf s_2}N-q\right)
\end{equation}
{\em This is the probability that two stationary points uniformly drawn from the ensemble
of all stationary points with fixed $E$ and $\mu^*$ happen to be at overlap $q$.}
Though these are evaluated for a given energy, index, etc, we shall omit these
subindices for simplicity.

The moments of this distribution $q^{(p)}$ are given by
\begin{equation}
  \begin{aligned}
    q^{(p)}
    &\equiv\int_0^1dq\,q^pP(q)
    =\frac1{N^p}\sum_{i_1\cdots i_p}\langle s_{i_1}\cdots s_{i_p}\rangle\langle s_{i_1}\cdots s_{i_p}\rangle
    =\frac1{N^p} \; \frac{1}{{\cal{N}}^2} \left\{ \sum_{{\mathbf s}_1,{\mathbf s}_2}\; \sum_{i_1\cdots i_p} s^1_{i_1}\cdots s^1_{i_p} s^2_{i_1}\cdots s^2_{i_p}\right\} \\
    &=\frac{1}{{\mathcal{N}}^2} \left\{\sum_{{\mathbf s}_1,{\mathbf s}_2}\left(\frac{\mathbf s_1\cdot\mathbf s_2}N\right)^p\right\}
    =\lim_{n\to0} \left\{\sum_{{\mathbf s}_1,{\mathbf s}_2,\ldots,\mathbf s_n} \left(\frac{\mathbf s_1\cdot\mathbf s_2}N\right)^p\right\}
  \end{aligned}
\end{equation}
The $(n-2)$ extra replicas provide the normalization, with
$\lim_{n\to0}\mathcal N^{n-2}=\mathcal N^{-2}$.  Replacing the sums over
stationary points with integrals over the Kac--Rice measure, the average over
disorder (again, for fixed energy and index) gives
\begin{equation}
  \begin{aligned}
    \overline{q^{(p)}} &=\overline{\frac1{N^p}\sum_{i_1\cdots i_p}\langle s_{i_1}\cdots s_{i_p}\rangle\langle s_{i_1}\cdots s_{i_p}\rangle}
    =\lim_{n\to0}{\int\overline{\prod_a^n d\nu(\mathbf s_a)}\,\left(\frac{\mathbf s_1\cdot\mathbf s_2}N\right)^p} \\
    &=\lim_{n\to0}{\int D[C,R,D] \,
    \left(C_{12}\right)^p\; } e^{nN\Sigma[C,R,D]}
    =\lim_{n\to0}{\int D[C,R,D] \,\frac1{n(n-1)}\sum_{a\neq b}\left(C_{ab}\right)^p\; } e^{nN\Sigma[C,R,D]}
  \end{aligned}
\end{equation}
In the last line, we have used that there is nothing special about replicas one and two.
Using the Parisi ansatz, evaluating by saddle point  {\em summing over all the $n(n-1)$ saddles related by permutation} we then have
\begin{equation}
  \overline{q^{(p)}}=\int_0^1 dx\,c^p(x) = \int_0^1 dq\,q^p P(q) \qquad \qquad {\mbox{concluding}} \qquad P(q)=\frac{dx}{dq}=\left(\frac{dc}{dx}\right)^{-1}\bigg|_{c(x)=q}
\end{equation}
The appeal of Parisi to properties of pure states is unnecessary here, since
the stationary points are points.

With this
established, we now address what it means for $C$ to have a nontrivial
replica-symmetry broken structure. When $C$ is replica symmetric, drawing two
stationary points at random will always lead to the same overlap. In the case
when there is no linear field and $q_0=0$, they will always have overlap zero,
because the second point will almost certainly lie on the equator of the sphere
with respect to the first. Though other stationary points exist nearby the
first one, they are exponentially fewer and so will be picked with vanishing
probability in the thermodynamic limit.

\begin{figure}
  \centering
  \includegraphics{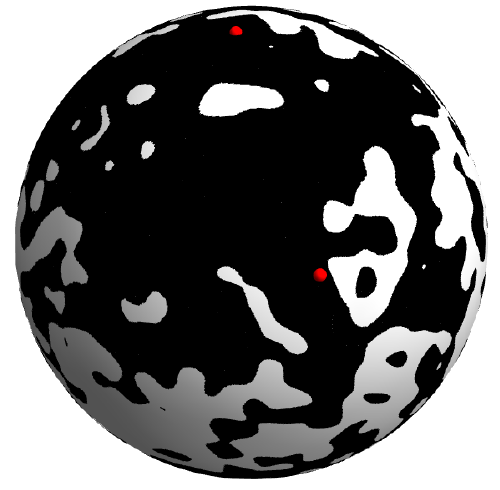}
  \hfill
  \includegraphics{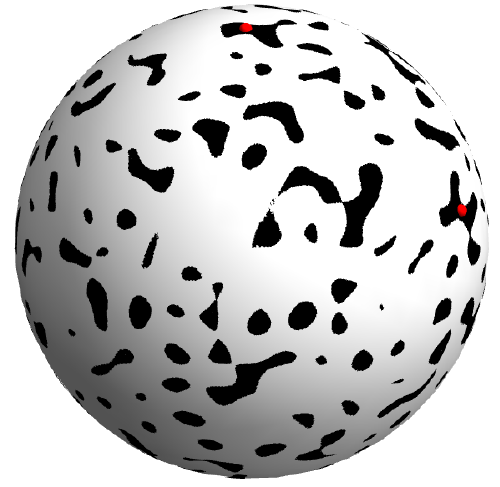}
  \hfill
  \includegraphics{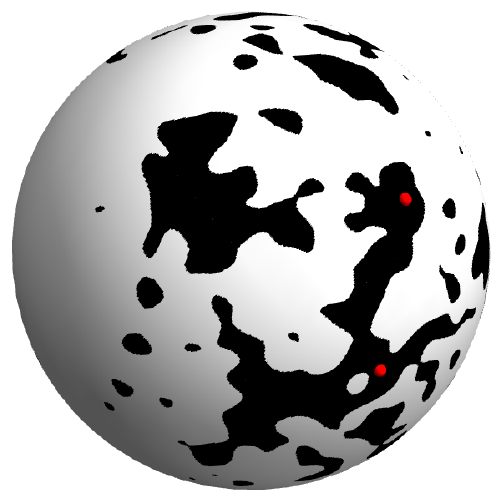}

  \caption{
    A cartoon visualizing how to interpret replica symmetry breaking solutions
    in the complexity. The black region show schematically areas where
    stationary points of a given energy can be found. Left: When the region
    is connected, pairs of stationary points exist at any overlap, but the
    vast majority of pairs are orthogonal. Center: When there are exponentially
    many disconnected regions of similar size, the vast majority of pairs will
    be found in different, orthogonal regions. Right: When there are a few
    large disconnected regions, pairs have a comparable probability to be found
    in different regions or in the same region. This gives rise to two (or
    more) possible overlaps.
  } \label{fig:cartoon}
\end{figure}

When $C$ is replica-symmetry broken, there is a nonzero probability of picking
a second stationary point at some other overlap. This can be interpreted by
imagining the level sets of the Hamiltonian in this scenario. If the level sets
are disconnected but there are exponentially many of them distributed on the
sphere, one will still find zero average overlap. However, if the disconnected
level sets are \emph{few}, i.e., less than order $N$, then it is possible to
draw two stationary points from the same set with nonzero probability.
Therefore, the picture in this case is of few, large basins each containing
exponentially many stationary points. A cartoon of this picture is shown in Fig.~\ref{fig:cartoon}.

\subsubsection{A tractable example}

One can construct a schematic 2RSB model from two 1RSB models.
Consider two independent pure models of size $N$ and with $p_1$-spin and $p_2$-spin couplings,
respectively, with energies $H_{p_1}({\mathbf s})$ and
$H_{p_2}({\boldsymbol\sigma})$, and couple them weakly with
$\varepsilon \; {\boldsymbol \sigma} \cdot {\mathbf s}$. The landscape of the
pure models is much simpler than that of the mixed because, in these models,
fixing the stability $\mu$ is equivalent to fixing the energy: $\mu=pE$. This
implies that at each energy level there is only one type of stationary point.
Therefore, for the pure models our formulas for the complexity and its Legendre
transforms are functions of one variable only, $E$, and each instance of
$\mu^*$ inside must be replaced with $pE$.

In the joint model, we wish to fix the total energy, not the energies of the
individual two models. Therefore, we insert a $\delta$-function containing
$(E_1+E_2)-E$ and integrate over $E_1$ and $E_2$. This results in a joint
complexity (and Legendre transform)
\begin{align}
    e^{N\Sigma(E)}&=\int dE_1\, dE_2\, d\lambda \,\exp\left\{N\left[\Sigma_1(E_1) + \Sigma_2(E_2) + O(\varepsilon) -\lambda\Big((E_1+E_2)-E\Big)\right]\right\} \\
    e^{NG(\hat \beta)}&=\int dE\, dE_1\, dE_2\, d\lambda\, \exp\left\{N\left[-\hat\beta E+\Sigma_1(E_1) + \Sigma_2(E_2) + O(\varepsilon) -\lambda\Big((E_1+E_2)-E\Big)\right]\right\}
\end{align}
The saddle point is given by $\Sigma_1'(E_1)=\Sigma_2'(E_2)=\hat\beta$, provided that
both $\Sigma_1(E_1)$ and $\Sigma_2(E_2)$ are non-zero. In this situation, two systems are `thermalized',
and, because many points contribute, the overlap between two global
configurations is zero:
\begin{equation}
  \frac1 {2N}\Big\langle({\mathbf s^1},{\boldsymbol\sigma^1})\cdot ({\mathbf s^2},{\boldsymbol\sigma^2})\Big\rangle
  =\frac1 {2N}\Big[\langle{\mathbf s^1}\cdot {\mathbf s^2}\rangle+ \langle{\boldsymbol\sigma^1}\cdot {\boldsymbol\sigma^2}\rangle\Big]
  =0
\end{equation}
This is the `annealed' phase of a Kac-Rice calculation.

Now start going down in energy, or up in $\hat \beta$: there will be a point
$E_c$ or $\hat \beta_c$ at which one of the subsystems (say it is system one)
freezes at its lowest energy density, while system two is not yet frozen. At
this point, $\Sigma_1(E_1)=0$ and $E_1$ is the ground state energy. At an even
higher value $\hat \beta=\hat \beta_f$, both systems will become frozen in
their ground states.  For $\hat \beta_f > \hat \beta> \hat \beta_c$ one system
is unfrozen, while the other is, because of coupling, frozen at inverse
temperature $\hat \beta_c$.  The overlap between two solutions in this intermediate phase is
\begin{equation}
  \frac1 {2N}\Big\langle({\mathbf s^1},{\boldsymbol\sigma^1})\cdot ({\mathbf s^2},{\boldsymbol\sigma^2})\Big\rangle
  =\frac1 {2N}\Big[\langle{\mathbf s^1}\cdot {\mathbf s^2}\rangle+ \langle{\boldsymbol\sigma^1}\cdot {\boldsymbol\sigma^2}\rangle\Big]
  = \frac1 {2N}\langle{\mathbf s^1}\cdot {\mathbf s^2}\rangle>0
\end{equation}
which is nonzero because there are only a few low-energy stationary points in system one, and
there is a nonvanishing probability of selecting one of them twice.
The distribution of this overlap is one-half the overlap distribution of a
frozen spin-glass at temperature $\hat \beta$, a 1RSB system like the Random
Energy Model. The value of $x$ corresponding to it depends on $\hat \beta$,
starting at $x=1$ at $\hat \beta_c$ and decreasing with increasing $\hat
\beta$.  Globally, the joint complexity of the system is  1RSB, but note that the
global overlap between different states is at most $1/2$.  At $\hat \beta>\hat
\beta_f$ there is a further transition.

This schematic example provides a metaphor for considering what happens in
ordinary models when replica symmetry is broken. At some point certain degrees
of freedom `freeze' onto a subextensive number of possible states, while the
remainder are effectively unconstrained. The overlap measures something in the
competition between the number of these unconstrained subregions and their
size.

\subsection{\textit{R} and \textit{D}: response functions}

The matrix field $R$ is related to responses of the stationary points to
perturbations of the tensors $J$. One adds to the Hamiltonian a random term
$\varepsilon_p \tilde H_p = -\frac1{p!}\varepsilon_p \sum_{i_1\cdots i_p} \tilde J_{i_1\cdots i_p}
s_{i_1}\cdots s_{i_p}$, where the $\tilde J$ are random Gaussian uncorrelated with
the $J$s and having variance $\overline{\tilde J^2}=p!/2N^{p-1}$.  The response to these is:
\begin{equation}
\frac1N\overline{\frac{\partial \langle \tilde H_p \rangle  }  {\partial \varepsilon_p}}
     =\lim_{n\to0}\int\left(\prod_a^nd\nu(\mathbf s_a)\right)\sum_b^n\left[
      \hat\beta\left(\frac{\mathbf s_1\cdot\mathbf s_b}N\right)^p+
        p\left(-i\frac{\mathbf s_1\cdot\hat{\mathbf s}_b}N\right)\left(\frac{\mathbf s_1\cdot\mathbf s_b}N\right)^{p-1}
      \right]
\end{equation}
Taking the average of this expression over disorder and averaging over the equivalent replicas in the integral gives, similar to before,
\begin{equation}
  \begin{aligned}
  \frac1N\overline{\frac{\partial\langle \tilde H_p \rangle  }  {\partial \varepsilon_p}}
  %  \overline{\frac1{N^p}\sum_{i_1\cdots i_p}\frac{\partial\langle s_{i_1}\cdots s_{i_p}\rangle}{\partial J^{(p)}_{i_1\cdots i_p}}}
    &=\lim_{n\to0}\int D[C,R,D]\,\frac1n\sum_{ab}^n(\hat\beta C_{ab}^p+pR_{ab}C_{ab}^{p-1})e^{nN\Sigma[C,R,D]}\\
    &=\hat\beta+pr_d-\int_0^1dx\,c^{p-1}(x)(\hat\beta c(x)+pr(x))
  \end{aligned}
\end{equation}
The responses as defined by this average perturbation in the pure $p$-spin
energy can be directly related to responses in the tensor polarization of the
stationary points:
\begin{equation}
  \frac1{N^p}\overline{\sum_{i_1\cdots i_p}\frac{\partial\langle s_{i_1}\cdots s_{i_p}\rangle}{\partial J^{(p)}_{i_1\cdots i_p}}}
  =\frac1N\overline{\frac{\partial\langle \tilde H_p \rangle  }  {\partial \varepsilon_p}}
\end{equation}
In particular, when the energy is unconstrained ($\hat\beta=0$) and there is replica symmetry, the above formulas imply that
\begin{equation}
  \frac1N\sum_i\frac{\partial\langle s_i\rangle}{\partial J_i^{(1)}}
  =r_d
\end{equation}
i.e., adding a linear field causes a response in the average stationary point
location proportional to $r_d$. If positive, for instance, stationary points
tend to align with a field. The energy constraint has a significant
contribution due to the perturbation causing stationary points to move up or
down in energy.

The matrix field $D$ is related to the response of the complexity to
perturbations of the variance of the tensors $J$. This can be found by taking
the expression for the complexity and inserting the dependence of $f$ on the
coefficients $a_p$, then differentiating:
\begin{equation}
  \begin{aligned}
    \frac{\partial\Sigma}{\partial a_p}
    =\frac14\lim_{n\to0}\frac1n\sum_{ab}^n\left[
      \hat\beta^2C_{ab}^p+p(2\hat\beta R_{ab}-D_{ab})C_{ab}^{p-1}+p(p-1)R_{ab}^2C_{ab}^{p-2}
    \right]
  \end{aligned}
\end{equation}
In particular, when the energy is unconstrained ($\hat\beta=0$) and there is no replica symmetry breaking,
\begin{equation}
  \frac{\partial\Sigma}{\partial a_1}=-\frac14\lim_{n\to0}\frac1n\sum_{ab}D_{ab}=-\frac14d_d
\end{equation}
i.e., adding a random linear field decreases the complexity of solutions by an amount proportional to $d_d$ in the variance of the field.

When the saddle point of the Kac--Rice problem is supersymmetric,
\begin{equation}
  \frac{\partial\Sigma}{\partial a_p}
  =\frac{\hat\beta}4\overline{\frac1{N^p}\sum_{i_1\cdots i_p}\frac{\partial\langle s_{i_1}\cdots s_{i_p}\rangle}{\partial J^{(p)}_{i_1\cdots i_p}}}+\lim_{n\to0}\frac1n\sum_{ab}^np(p-1)R_{ab}^2C_{ab}^{p-2}
\end{equation}
and in particular for $p=1$
\begin{equation}
  \frac{\partial\Sigma}{\partial a_1}
  =\frac{\hat\beta}4\overline{\frac1N\sum_i\frac{\partial\langle s_i\rangle}{\partial J_i^{(1)}}}
\end{equation}
i.e., the change in complexity due to a linear field is directly related to the
resulting magnetization of the stationary points for supersymmetric minima.

\section{Conclusion}
\label{se:conclusion}

We have constructed a  replica solution for the general  problem of finding
saddles of random mean-field landscapes, including systems with many steps of
RSB.  For systems with full RSB, we find that minima are exponentially
subdominant with respect to saddles at all energy densities above the ground
state.  The solution should be subjected to standard checks, like the
examination of its stability with respect to other RSB schemes. The solution
contains valuable geometric information that has yet to be extracted in all
detail, for example considering several copies of the system
\cite{Cavagna_1997_Structure}, or the extension to complex variables
\cite{Kent-Dobias_2021_Complex, Kent-Dobias_2022_Analytic}.

A first and very important application of the method here is to perform the
calculation for high dimensional spheres, where it would give us a clear
understanding of what happens in realistic low-temperature jamming dynamics
\cite{Maimbourg_2016_Solution}. More simply, examining the landscape of a
spherical model with a glass to glass transition from 1RSB to RS, like the
$2+4$ model when $a_4$ is larger than we have taken it in our example, might
give insight into the cases of interest for Gardner physics
\cite{Crisanti_2004_Spherical, Crisanti_2006_Spherical}. In any case, our
analysis of typical 1RSB and FRSB landscapes indicates that the highest energy
signature of RSB phases is in the overlap structure of the
high-index saddle points. Though measuring the statistics of saddle points is
difficult to imagine for experiments, this insight could find application in
simulations of glass formers, where saddle-finding methods are possible.

A second application is to evaluate in more detail the landscape of these RSB
systems. In particular, examining the complexity of stationary points with
non-extensive indices (like rank-one saddles), the complexity of pairs of
stationary points at fixed overlap, or the complexity of energy barriers
\cite{Auffinger_2012_Random, Ros_2019_Complexity}. These other properties of
the landscape might shed light on the relationship between landscape RSB and
dynamical features, like the algorithmic energy $E_\mathrm{alg}$, or the asymptotic level reached by physical dynamics. For our 1RSB
example, because $E_\mathrm{alg}$ is just below the energy where
dominant saddles transition to a RSB complexity, we speculate that
$E_\mathrm{alg}$ may be related to the statistics of minima connected to the
saddles at this transition point.

\begin{appendix}

  \section{Hierarchical matrix dictionary}
  \label{sec:dict}

  Each row of a hierarchical matrix is the same up to permutation of their
  elements.  The so-called $k$RSB ansatz has $k+2$ different values in each
  row. If $A$ is an $n\times n$ hierarchical matrix, then $n-x_1$ of those
  entries are $a_0$, $x_1-x_2$ of those entries are $a_1$, and so on until
  $x_k-1$ entries of $a_k$, and one entry of $a_d$, corresponding to the
  diagonal.  Given such a matrix, there are standard ways of producing the sum
  and determinant that appear in the free energy. These formulas are, for an
  arbitrary $k$RSB matrix $A$ with $a_d$ on its diagonal (recall $q_d=1$),
  \begin{equation} \label{eq:replica.sum}
    \lim_{n\to0}\frac1n\sum_{ab}^nA_{ab}
    =a_d-\sum_{i=0}^k(x_{i+1}-x_i)a_i
  \end{equation}
  \begin{equation} \label{eq:replica.logdet}
    \begin{aligned}
      \lim_{n\to0}\frac1n\ln\det A
      &=
      \frac{a_0}{a_d-\sum_{i=0}^k(x_{i+1}-x_i)a_i}
      +\frac1{x_1}\log\left[
        a_d-\sum_{i=0}^{k}(x_{i+1}-x_i)a_i
      \right]\\
      &\hspace{10pc}-\sum_{j=1}^k(x_j^{-1}-x_{j+1}^{-1})\log\left[
          a_d-\sum_{i=j}^{k}(x_{i+1}-x_i)a_i-x_ja_j
      \right]
    \end{aligned}
  \end{equation}
  where $x_0=0$ and $x_{k+1}=1$. The sum of two hierarchical matrices results
  in the sum of each of their elements: $(a+b)_d=a_d+b_d$ and
  $(a+b)_i=a_i+b_i$.  The product $AB$ of two hierarchical matrices $A$
  and $B$ is given by
  \begin{align} \label{eq:replica.prod}
    (a\ast b)_d&=a_db_d-\sum_{j=0}^k(x_{j+1}-x_j)a_jb_j \\
    (a\ast b)_i&=b_da_i+a_db_i-\sum_{j=0}^{i-1}(x_{j+1}-x_j)a_jb_j+(2x_{i+1}-x_i)a_ib_i
    -\sum_{j=i+1}^k(x_{j+1}-x_j)(a_ib_j+a_jb_i)
  \end{align}

  There is a canonical mapping between the parameterization of a hierarchical
  matrix described above and a functional parameterization that is particularly
  convenient in the twin limit $n\to0$ and $k\to\infty$
  \cite{Parisi_1980_Magnetic, Mezard_1991_Replica}. The distribution of
  diagonal elements of a matrix $A$ is parameterized by a continuous function
  $a(x)$ on the interval $[0,1]$, while its diagonal is still called $a_d$.
  Define for any function $g$ the average
  \begin{equation}
    \langle g\rangle=\int_0^1dx\,g(x)
  \end{equation}
  The sum of two hierarchical matrices so parameterized results in the  sum of
  these functions. The product $AB$ of hierarchical matrices $A$ and $B$ gives
  \begin{align} \label{eq:cont.replica.prod}
    (a\ast b)_d&=a_db_d-\langle ab\rangle \\
    (a\ast b)(x)&=(b_d-\langle b\rangle)a(x)+(a_d-\langle a\rangle)b(x)
    -\int_0^xdy\,\big(
      a(x)-a(y)
      \big)\big(
      b(x)-b(y)
    \big)
  \end{align}
  The sum over all elements of a hierarchical matrix $A$ gives
  \begin{equation}
    \lim_{n\to0}\frac1n\sum_{ab}A_{ab}=a_d-\langle a\rangle
  \end{equation}
  The $\ln\det=\operatorname{Tr}\ln$ becomes
  \begin{equation} \label{eq:replica.det.cont}
    \lim_{n\to0}\frac1n\ln\det A
    =\ln(a_d-\langle a\rangle)
    +\frac{a(0)}{a_d-\langle a\rangle}
    -\int_0^1\frac{dx}{x^2}\ln\left(\frac{a_d-\langle a\rangle-xa(x)+\int_0^xdy\,a(y)}{a_d-\langle a\rangle}\right)
  \end{equation}
\end{appendix}

\paragraph{Acknowledgements}
The authors would like to thank Valentina Ros for helpful discussions.

\paragraph{Funding information}
JK-D and JK are supported by the Simons Foundation Grant No.~454943.

\printbibliography

\end{document}